\documentclass{aa}  
\usepackage{txfonts}

\usepackage{natbib}
\usepackage{graphicx}
\usepackage{color}

\newcommand{\carcsec}{$\mbox{.\hspace{-0.5ex}}^{\prime\prime}$}
\newcommand{\sunrise}{\textsc{Sunrise}}%

\graphicspath{{images_kahil_ms_red/}}
\begin{document}

   \title{Intensity contrast of solar plage as a function of magnetic flux at high spatial resolution }

   \author{F. Kahil,
          \inst{1}
          T.L. Riethm\"{u}ller,\inst{1}
          \and
          S.K. Solanki\inst{1}\inst{2}
          }

   \institute{Max-Planck-Institut f\"{u}r Sonnensystemforschung, Justus-von-Liebig-weg 3,	37077 G\"{o}ttingen, Germany\\
              \email{kahil@mps.mpg.de}
         \and
             School of Space Research, Kyung Hee University, Yongin, Gyeonggi 446-701, Republic of Korea\\
             }


 
  \abstract
 { Magnetic elements have an intensity contrast that depends on the type of region they are located in (e.g. quiet Sun, or active region plage). Observed values also depend on the spatial resolution of the data. Here we investigate the contrast-magnetic field dependence in active region plage observed near disk center with $\sunrise$ during its second flight in 2013. The wavelengths under study range from the visible at 525\,nm to the near ultraviolet (NUV) at 300\,nm and 397\,nm. We use quasi-simultaneous spectropolarimetric and photometric data from the Imaging Magnetograph eXperiment (IMaX) and the Sunrise Filter Imager (SuFI), respectively. We find that in all wavelength bands, the contrast exhibits a qualitatively similar dependence on the line-of-sight magnetic field, $B_{\rm LOS}$, as found in the quiet Sun, with the exception of the continuum at 525\,nm. There, the contrast of plage magnetic elements peaks for intermediate values of $B_{\rm LOS}$ and decreases at higher field strengths. By comparison, the contrast of magnetic elements in the quiet Sun saturates at its maximum value at large $B_{\rm LOS}$. We find that the explanation of the turnover in contrast in terms of the effect of finite spatial resolution of the data is incorrect with the evidence provided by the high-spatial resolution $\sunrise$ data, as the plage magnetic elements are larger than the quiet Sun magnetic elements and are well-resolved. The turnover comes from the fact that the core pixels of these larger magnetic elements are darker than the quiet Sun. We find that plages reach lower contrast than the quiet Sun at disk center at wavelength bands formed deep in the photosphere, such as the visible continuum and the 300\,nm band. This difference decreases with formation height and disappears in the Ca\,{\sc ii} H core, in agreement with empirical models of magnetic element atmospheres.} 

   \keywords{Sun: activity - Sun: photosphere - Sun: chromosphere - Sun: faculae, plages, quiet Sun - Sun: magnetic field, spectropolarimetry, photometry
               }
\titlerunning{Intensity contrast of solar plage}
\authorrunning{Kahil et al.}
\maketitle

\section{Introduction}
In addition to quiet-Sun (QS) network, small-scale magnetic elements on the solar surface manifest themselves in the form of plage regions or faculae. 
Faculae are seen near the solar limb in white light as bright structures often surrounding sunspots and pores. The same structures seen in chromospheric emission are called plages. Here, we have used the latter nomenclature to describe such magnetic flux concentrations seen at wavelengths sampling heights ranging from the low photosphere to the lower chromosphere. The network is seen at the edges of supergranules, while plages are part of active regions (ARs). Network and plage both harbor kG magnetic fields \citep{stenflo_magnetic-field_1973}.

At the photospheric level, magnetic elements are located within the intergranular lanes and have been successfully described by flux tubes \citep{spruit_pressure_1976, solanki_small-scale_1993}. The magnetic field is swept to the dark intergranular lanes by granular motions and concentrated there by the magnetic flux expulsion mechanism \citep{parker_kinematical_1963}. The intensification of the magnetic field within these elements causes the internal gas pressure at a given height to drop to maintain total horizontal pressure equilibrium with the non-magnetic surroundings. This leads to an opacity depression, so that one can see into deeper layers of the photosphere within such elements. Their brightness is dependent on their size, the extent of the opacity depression (i.e., effectively the magnetic field strength), which together determine the effective heating from the surrounding hot walls, which correspond to the sides of the surrounding granules.

Small-scale magnetic elements are also linked to the heating of the chromopshere and corona. For the chromosphere, this is indicated by the correlation found between the brightness in the cores of Ca II H and K lines and the magnetic field \citep[]{skumanich_statistical_1975,schrijver_relations_1989,ortiz_how_2005,rezaei_relation_2007,loukitcheva_relationship_2009,kahil_brightness_2017}, while the importance of small magnetic features for coronal heating has been pointed out by \cite{zhang_lifetime_1998,ishikawa_comparison_2009, zhou_solar_2010,chitta_solar_2017}. Together, the network and faculae are responsible for the brightening of the Sun with higher magnetic activity on time scales longer than solar rotation \citep{krivova_reconstruction_2003,yeo_solar_2017}. 

The reported brightness of magnetic elements with respect to the quiet surrounding (i.e. their contrast) in the photosphere, along with their morphological structure depends on the wavelength band in which these structures are seen, their position on the solar disk, the selection method used to identify them, and the spatial resolution at which they are observed. Consequently, different observations sometimes gave seemingly contradictory results regarding their photometric and magnetic properties.
Below we give an overview of some of the studies that have looked into the contrast-magnetic field relationship, concentrating on the (near) disk center observations.\\

At a spatial resolution of 4$^{\prime\prime}$, \citet{ortiz_intensity_2002} used simultaneous full-disk magnetograms and continuum images (at 676.8 nm) acquired by the Michelson Doppler Imager (MDI) on-board the Solar and Heliospheric Observatory (SoHO). At disk center, the contrast of active-region faculae (identified by their higher magnetic field values) is negative while it is positive for smaller magnetic elements (the network). They attributed the contrast difference to the size of the features, whereby the radiative heating from the surrounding walls is more important and effective when the cross-section of a flux tube is smaller. This study was redone at higher resolution by \cite{yeo_intensity_2013} using HMI data. They obtained essentially the same qualitative results, with the differences that the contrast values at weak magnetogram signals in their identified network are negative.

\citet{frazier_multi-channel_1971} used simultaneous observations of photospheric magnetic field inferred from the Fe\,{\sc i} 5233.0\, \AA{} line and continuum filtergrams at Fe\,{\sc i} 5250.2\, \AA{} in addition to line-core data in Fe\,{\sc i} 5250.2\, \AA{} and Ca\,{\sc ii} K of two active regions at disk center. Whereas continuum contrast increased up until 200\,G followed by a decrease at higher fields, the line-core and Ca\,{\sc ii} K contrasts kept on increasing with increasing magnetogram signal. For the core of another photospheric line, Fe\,{\sc i} 6173\, \AA{}, \cite{yeo_intensity_2013} obtained that the line contrast remains positive at all studied field strengths, but it does decrease somewhat at higher field strengths. For Ca\,{\sc ii} H and K this qualitative behavior was confirmed by, for example, \cite{skumanich_statistical_1975}, \cite{schrijver_relations_1989}, \cite{loukitcheva_relationship_2009}, \cite{kahil_brightness_2017}, although the exact dependence varied. 

\citet{title_differences_1992}, \citet{topka_properties_1992} and \citet{lawrence_contrast_1993} used AR data from the 50\,cm Swedish Solar Vacuum Telescope (SVST), while \citet{lawrence_contrast_1993} also analyzed the quiet-Sun network.
At disk center, and at a spatial resolution of 0.3$^{\prime\prime}$ their measured continuum brightness at visible wavelengths (676.8, 525, 557.6, and 630.2 nm) is negative (less than the mean QS intensity) for all magnetogram signals. The line core brightness at 676.8 nm increases until 600\,G, then drops monotonically at higher fields in active regions.
They reported higher (more positive) contrast in the quiet Sun \citep{lawrence_contrast_1993}.

\citet{kobel_continuum_2011} looked into the brightness relationship with the magnetogram signal of both, QS and AR plages. They used spectropolarimetric data at 630.2\,nm  from the Solar Optical Telecope on-board Hinode (spatial resolution of 0.3$^{\prime\prime}$). At disk center, the contrast of both regions showed the same behaviour when plotted against the longitudinal field strength: contrast peaking at 700\,G, and turning over at higher fields. This partly contradicts the results of \cite{title_differences_1992}; \cite{topka_properties_1992} and \cite{lawrence_contrast_1993}. \citet{kobel_continuum_2011} also obtained higher contrast in the QS compared to AR plages, which confirmed the results of \cite{title_differences_1992} and \cite{lawrence_contrast_1993}.

At a spatial resolution of 0.1$^{\prime\prime}$, \citet{berger_contrast_2007} studied the relationship between the G-band contrast of solar magnetic elements and the magnetic field at different heliocentric angles using the Swedich Solar Telescope \citep[SST,][]{scharmer_1-meter_2003}. For faculae close to disk center ($\mu=0.97$), the G-band contrast is positive for fields up until 700\,G and becomes negative for higher fields, with sunspots and pores excluded from their study.
Also using the SST, \citet{narayan_small-scale_2010} analysed the dynamical properties of solar magnetic elements close to disk center, using spectropolarimetric data collected by the CRisp Imaging SPectropolarimter (CRISP) in the Fe\,{\sc i} line at 630.25\, nm, with the magnetic field obtained via Milne-Eddington inversions. After masking the dark pores in their images, their scatterplot of the continuum contrast against the longitudinal component of the magnetic field showed an initial decrease of the contrast for fields less than 200\, G, then an increase until 600\,G followed by a drop  below the QS reference for higher fields.

So far the observational results show a strong disagreement with the predictions of radiative magnetohydrodynamic simulations (MHD) at their original resolution. According to such simulations, in agreement with  the flux tube model, the continuum contrast of small-scale magnetic elements should increase monotonically with the magnetic field strength \citep[][their Fig.~13]{vogler_simulations_2005}. In all observational studies the continuum contrast (in visible wavelengths) of quiet-Sun network and faculae decreases at higher fields, even when excluding pores in the case of faculae. This decrease was attributed to the limited spatial resolution of observations (even if they are carried out using space-borne instruments) compared to the higher resolution of MHD simulations \citep[]{rohrbein_is_2011,danilovic_relation_2013}. At the spatial resolution of 0.15$^{\prime\prime}$ achieved by the Imaging Magnetograph eXperiment (IMaX) onboard $\sunrise$, \cite{kahil_brightness_2017} confirmed this hypothesis by demonstrating that the continuum contrast at 525\,nm of a quiet-Sun region saturated at the strongest fields. Inspecting this relationship in the larger plage features at the spatial resolution of IMaX is one of the primary aims of the current work. \\

A qualitative and quantitative study has been carried out by \cite{kahil_brightness_2017} (hereafter paper I) for the pixel-by-pixel relationship of contrast to magnetic field in quiet-Sun network at disk center. We used observations recorded by the balloon-borne observatory $\sunrise$ during its first flight in 2009 ($\sunrise$ I), in the UV and visible wavelength ranges. Here, we look into the same relationship, but for plage regions observed by $\sunrise$ during its second flight in 2013 ($\sunrise$ II).  
The data and analysis techniques are described in Section \ref{observations}. In Section \ref{results} we show our results and compare them to similar studies and to those obtained in paper I for quiet-Sun data. In Section~\ref{conclusion} we discuss our findings and come up with conclusions and insights for future investigations.

\section{Observations and data reduction}
\label{observations}
The time series used in this study is recorded with the 1-m telescope on the balloon-borne observatory $\sunrise$ \citep{solanki_sunrise:_2010,barthol_sunrise_2011,berkefeld_wave-front_2011} during its second flight on 2013 June 12 \citep{solanki_second_2017}. We use simultaneous spectropolarimetric and imaging data collected by the two scientific instruments onboard, an imaging spectropolarimeter \citep[IMaX;][]{martinezpillet_imaging_2011} and a UV filter imager \citep[SuFI;][]{gandorfer_filter_2011}. The observations were made close to disk center (cosine of heliocentric angle $\mu$ = 0.93), targeting the active region (AR) NOAA 11768 (see Section~\ref{inversions} for more details on the AR field of view). 
\subsection{IMaX data}
\label{imax}
The spectropolarimetric data were acquired by the Imaging Magnetograph eXperiment \citep[IMaX;][]{martinezpillet_imaging_2011} between 23:39 and 23:55 UT in their V8-4 mode. This mode consists of measuring the full Stokes vector $(I,Q,U,V)$ at 8 wavelength positions around the center of the photospheric Fe\,{\sc i} 5250.2\, \AA{} line, with four accumulations at each wavelength. The spectral positions sampled by IMaX were located at $\Delta \lambda$ = $-120$, $-80$, $-40$, $0$, $+40$, $+80$, $+120$\, m\AA{} from the line center of rest wavelength $\lambda_0$ = 5250.2\, \AA{} with an additional one in the red continuum (at $\Delta \lambda$ = $+227$\, m\AA{}).
With an exposure time of $250$ ms for each of the 4 polarization measurements at each of the eight spectral positions, the total cadence achieved for this mode was $36.5$ s. The 17 minutes time series consists of 28 sets of observations, i.e. images at 8 wavelengths in all 4 Stokes profiles, each image covering $51^{\prime\prime}\times51^{\prime\prime}$ on the solar surface with a platescale of 0\carcsec{}0545 pixel$^{-1}$.

During the flight, images were stabilized, by the use of a tip-tilt mirror controlled by the Correlating-Wave Front Sensor \citep[CWS;][]{berkefeld_wave-front_2011}.
Since there is a time lag in the acquisition of the polarization states between two successive wavelength scans, they were interpolated with respect to time to compensate for the solar evolution time during the acquisition of each observation cycle. In addition, images were corrected for instrumental polarization, and for low-order wavefront aberrations to improve the spatial resolution \citep{martinezpillet_imaging_2011}. At this stage, data are called phase-diversity (PD) reconstructed (level~$2.2$), with a noise level of 7$\times$10$^{-3} I_c$ in Stokes~$V$ ($I_c$ is the Stokes~$I$ continuum intensity) and a spatial resolution of $0.15^{\prime\prime}-0.18^{\prime\prime}$. 

Finally, a very simple approach was taken to correct for the instrumental straylight by subtracting 25\% of the spatial mean Stokes~$I$ profiles from the individual profiles at each of the 8 sampled wavelengths. The Stokes~$Q,U$ and $V$ data were not corrected for straylight since it is considered to be non-polarized (see \cite{riethmuller_new_2017} for more details on the stray-light correction). After stray-light correction, data are called level~$2.3$ data.

\subsection{Inversions}
\label{inversions}
After applying the above corrections, IMaX data were inverted to retrieve the physical parameters needed for our study. The Stokes inversion code SPINOR \citep{frutiger_properties_2000}, which uses the STOPRO routines for the radiative transfer \citep{solanki_photospheric_1987} was employed. The main retrieved parameters of relevance for this study are the magnetic field strength ($B$), the line-of-sight velocity ($V_{\rm LOS}$), and field inclination ($\gamma$). 
A simple one-component atmospheric model was considered, with three optical depth nodes at $\log\tau=-2.5, -0.9, 0$ for the temperature and a height-independent magnetic field vector, line-of-sight velocity and micro-turbulence. More details on the inversion strategy can be found in \cite{solanki_second_2017} and \cite{kahil_brightness_2017}. The line-of-sight velocity maps obtained from the inversions were corrected at each pixel for the wavelength blueshift caused by the Etalon used for the spectral analysis, and a constant velocity (0.6\, km/s) was removed so that the spatially averaged velocity across the FOV is zero.

\begin{figure*}
\centering
\includegraphics[width=\textwidth]{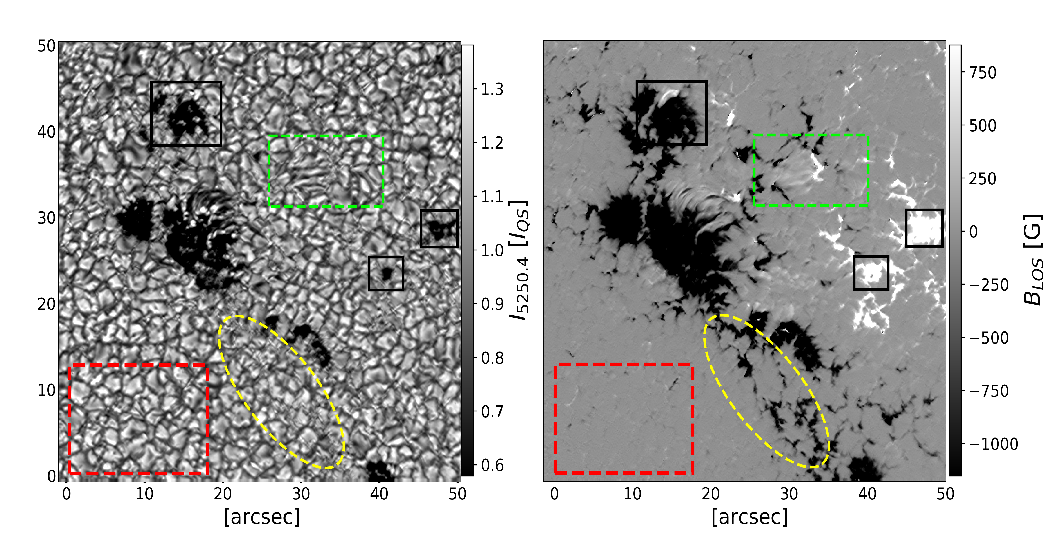}
\caption{\textit{Left panel:}
Continuum contrast at 525 nm. \textit{Right panel:} its co-spatial and co-temporal longitudinal magnetic field map retrieved from the inversions. The black boxes (solid lines) contain small pores characterized by contrast below unity and kG magnetic fields. The dashed green box encloses an area with emerging flux. The dashed yellow ellipse outlines a region of plage composed of magnetic elements embedded in intergranular lanes. The dashed red box contains a quiet-Sun internetwork region with weak fields (average of 20\,G) and mean contrast of unity (by definition).}
\label{fig1}

\end{figure*}

Figure~\ref{fig1} shows an example IMaX continuum map (at $\Delta \lambda=+227$\, m\AA{} from $\lambda_0$, left panel) and the corresponding magnetogram (longitudinal magnetic field, $B_{\rm LOS}$ map) returned by the inversions. Figure~\ref{fig1} displays a photospheric zoo of different structures with various photometric  and magnetic properties: a large pore, which has some penumbral structure attached to it, smaller pores with kG magnetic fields (black boxes), a flux emergence (FE) region (enclosed in the dashed green box), a small QS internetwork region (dashed red box) and plage (partly enclosed by the dashed yellow ellipse). The latter is located mainly in the vicinity of the pores and shows elongated structures or `ribbons' in the intergranular lanes as described in \citet{berger_solar_2004}. Such structures (wherever present in the FOV; i.e. not just in the yellow ellipse) will be analyzed in Section~\ref{turnover}.

\subsection{SuFI}
\label{sufi}

Simultaneously with IMaX, the Sunrise Filter Imager \citep[SuFI;][]{gandorfer_filter_2011} acquired high resolution images sampling the low-mid photosphere (300\,nm, bandwidth of 4.4\,nm) and the lower-chromosphere (397\,nm; core of Ca\,{\sc ii} H, bandwidth of 0.11\,nm) of the same region but with a narrower FOV of $15^{\prime\prime}\times38^{\prime\prime}$. The SuFI FOV covered a part of the leading polarity flux and flux-emergence regions, in addition to small pores and plage regions (see yellow and red boxes overlaid on IMaX FOV in Figs.~\ref{fig2}c and d). The integration time for both 300\,nm and 397\,nm images is 500 ms, with a cadence of 7\,s, and the corresponding  plate scale is 0\carcsec{}02 pixel$^{-1}$.
The spatial correction for wave-front aberrations was done using the Multi-Frame Blind Deconvolution technique \citep[MFBD;][]{van_noort_solar_2005}. The spatial resolution achieved at 397 nm after reconstruction is approximately 70\,km.
The reconstructed data are then corrected for instrumental straylight using the solar limb profiles recorded during the $\sunrise$~I flight. A stray-light modulation transfer function (MTF), with which images are deconvolved, is obtained upon comparing the observed limb profiles to those in the literature at each of the observed wavelength bands (A. Feller et al. 2018, in preparation).

\subsection{Alignment}
\label{alignment}
\begin{figure*}
\centering
\includegraphics[width=\textwidth]{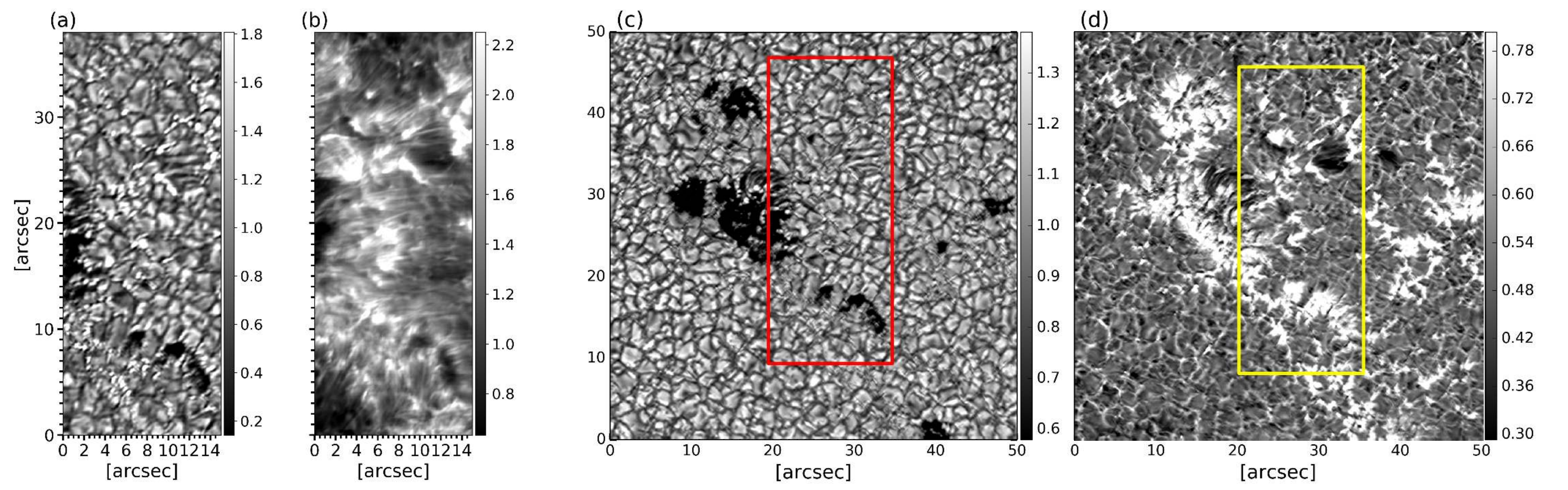}
\caption{The SuFI images coaligned to the IMaX FOV shown in Figure~\ref{fig1}. (a) SuFI 300 nm. (b) SuFI Ca\,{\sc ii} H. (c) IMaX continuum intensity (normalized to the mean quiet-Sun intensity) to which SuFI 300 nm (overlaid in red) is aligned. (d) IMaX line-core intensity derived from the inversions (see Section~\ref{contrast}) and normalized to the local continuum ($I_{+227}$), and to which calcium images (overlaid in yellow) are aligned. The yellow and red boxes are slightly misaligned due to the differential offsets in the SuFI wavelegnth channels. The gray scale is set to cover two times the rms range of each image.} 
\label{fig2}
\end{figure*}

Since we are comparing images recorded by SuFI with magnetograms obtained by IMaX, a precise alignment procedure has to be followed.
We apply here the same alignment procedure described in Paper~I for the quiet-Sun data, mainly resampling SuFI images to the pixel size of IMaX images, and using a cross-correlation technique to compute the shifts in both x and y directions. 
The SuFI images at 300 nm are aligned with IMaX Stokes~$I$ continuum images, since both of these spectral regions form in the low photosphere and show a normal granulation pattern (although bright points are more prominent at 300 nm, see Figure~\ref{fig2} and \cite{riethmuller_bright_2010}). The Ca\,{\sc ii} H data are aligned with IMaX Stokes~$I$ line-core images.
Both wavelength bands share similar spatial structures, such as the reversed granulation pattern (dark granules, bright intergranular lanes), a result of forming higher in the atmosphere (see Figs.~\ref{fig2}b and d). In the calcium images also other features besides reversed granulation are used (as the latter is best seen outside active regions). These include brightenings associated with strong magnetic fields, although these are somewhat diffuser in Ca\,{\sc ii} H  than in the line core of IMaX due to the expansion of the magnetic field, which forms a low-lying magnetic canopy in plage \citep{jafarzadeh_high-frequency_2017}. Note that since the IMaX line is weak, the IMaX line-core image shows a reversed granulation pattern only if normalized by the local continuum (see Fig.~\ref{fig2}d).

\subsection{Contrast}
\label{contrast}
At each pixel, the Stokes~$I$ continuum intensity is determined as the highest intensity of the synthetic best-fit profile retrieved from the inversion of IMaX data. We will consider these values instead of the ones observed at $\Delta \lambda = +227$\,m\AA{}, which can deviate from the true continuum if the neighboring line (Fe\,{\sc i} 5250.6\, \AA{}) influences the 227 m\AA{} intensity (due to a wavelength shift or broadening of the line).

We also refer to the inverted profiles to compute the IMaX line-core intensity as follows: we use the $V_{LOS}$ values returned by the inversions to derive the wavelength shift from the rest wavelength of the Fe\,{\sc i} line ($\lambda_0=5250.2$\, \AA{}) due to the Doppler effect. We add to it the shifts caused by the etalon blueshift and the convective blueshift that were already subtracted from the inverted $V_{LOS}$ maps, this gives the total shift of the line with respect to $\lambda_0$. The line-core intensity is the Stokes~$I$ value at the total computed wavelength shift. This approach was adopted since at pixels with higher magnetic fields, where Zeeman splitting is important, a simple Gaussian fit to Stokes~$I$ profiles was found to underestimate the line-core intensity. Moreover, the profiles fitted to IMaX data points by the inversions are far superior to Gaussian fits, since the latter do not take into account the various effects that contribute to the shape of the spectral line (temperature, density, magnetic field, etc..).

The relative intensity at each wavelength band was computed at every pixel by normalizing to the mean quiet-Sun intensity of a very quiet region (of weak magnetic fields). The contrast is the relative intensity minus unity, but we will use this terminology for the computed relative intensity for a consistent comparison with the literature. Hence the mean quiet Sun level will be attributed to a contrast of unity, brighter/darker pixels than the mean QS will have a contrast above/below unity. For the IMaX continuum and line core, the QS region is outlined by the dashed red rectangle in both panels of Figure~\ref{fig1}.

Due to the narrower field of view of SuFI, finding a quiet-Sun region in these data is more challenging, particularly as the QS region has to contain a large number of granules. Locating this region in the calcium images is even harder since one should look for dark regions with low fields, but due to the long fibrils found almost everywhere in the images (also protruding into neighbouring quiet regions), most of the low field regions are rather bright in calcium emission, affecting the computed mean quiet-Sun value. 

Luckily, quiet-Sun images are available as flatfields, which are sufficient for the purpose of obtaining the average quiet-Sun intensity. Obviously flatfields were recorded before and after the observations, but not during them. Fortunately, the photon flux at 397 nm does not depend on elevation angle (and hence time) so that the average quiet-Sun intensity does not vary with time and can be obtained straightforwardly from the flatfields. However, the 300 nm wavelength exhibits a cyclic variation due to absorption by the Earth's atmosphere. We use this day-to-night cycle variation to derive the mean quiet-Sun intensities at the times of our data acquisition.  In Appendix~\ref{appendix1} we provide more details on the computation of the mean quiet-Sun intensities in SuFI data.

\subsection{Masking}
\label{masking}
We are mainly interested in studying the relationship between contrast and magnetic field of magnetic elements in plages. These features are distributed everywhere in the FOV around the large and small pores. Hence, pores are masked by using an intensity threshold in the IMaX continuum maps. We compute the mean and standard deviation ($\sigma$) of each continuum image. After applying a low-pass filter consisting of the running mean over $33\times33$ pixels (to smooth the image), we discard pixels with intensities in the smoothed image lower than the mean minus $\sigma$ value. The smoothing greatly reduces the granular contrast. In particular intergranular lanes are much less dark, so that they lie above the threshold of mean minus $\sigma$ in the smoothed image. However, pores remain darker and the smoothing also leads to the exclusion of their immediate surroundings which are darkened by the pores after smoothing. To ignore pixels around the pores, which do not belong to magnetic elements (see Section~\ref{turnover} for a further discussion on the masking technique), the surroundings of pores in addition to the flux-emergence area (black and green boxes in Figure~\ref{fig1}) were masked manually. The flux-emergence region can be identified more easily when examining movies of Stokes~$I$ and $V$ images at the inner wavelength points, i.e.,  at $\Delta \lambda$ = $-40$, $0$, $+40$ m\AA{} from line center.

All unmasked pixels of the 28 images composing the IMaX time series are included in the pixel-by-pixel scatterplots analyzed in the next sections. For studying the contrast at 525 nm vs. $B_{\rm LOS}$, the pixels in the whole IMaX FOV of $51^{\prime\prime}\times51^{\prime\prime}$ are taken, while in the UV wavelengths we consider the pixels in the smaller FOV of $15^{\prime\prime}\times38^{\prime\prime}$ with the corresponding cropped $B_{\rm LOS}$ maps from the inverted IMaX data.
The masks computed from IMaX continuum maps were also used to mask the magnetic features in the corresponding IMaX magnetograms and SuFI images at 300 nm and 397 nm. 

\section{Results}
\label{results}

\subsection{Scatterplots of IMaX continuum and line-core constrasts vs. $B_{\rm LOS}$}
\label{results_imax}
\begin{figure}
\centering
   \includegraphics[width=\hsize]{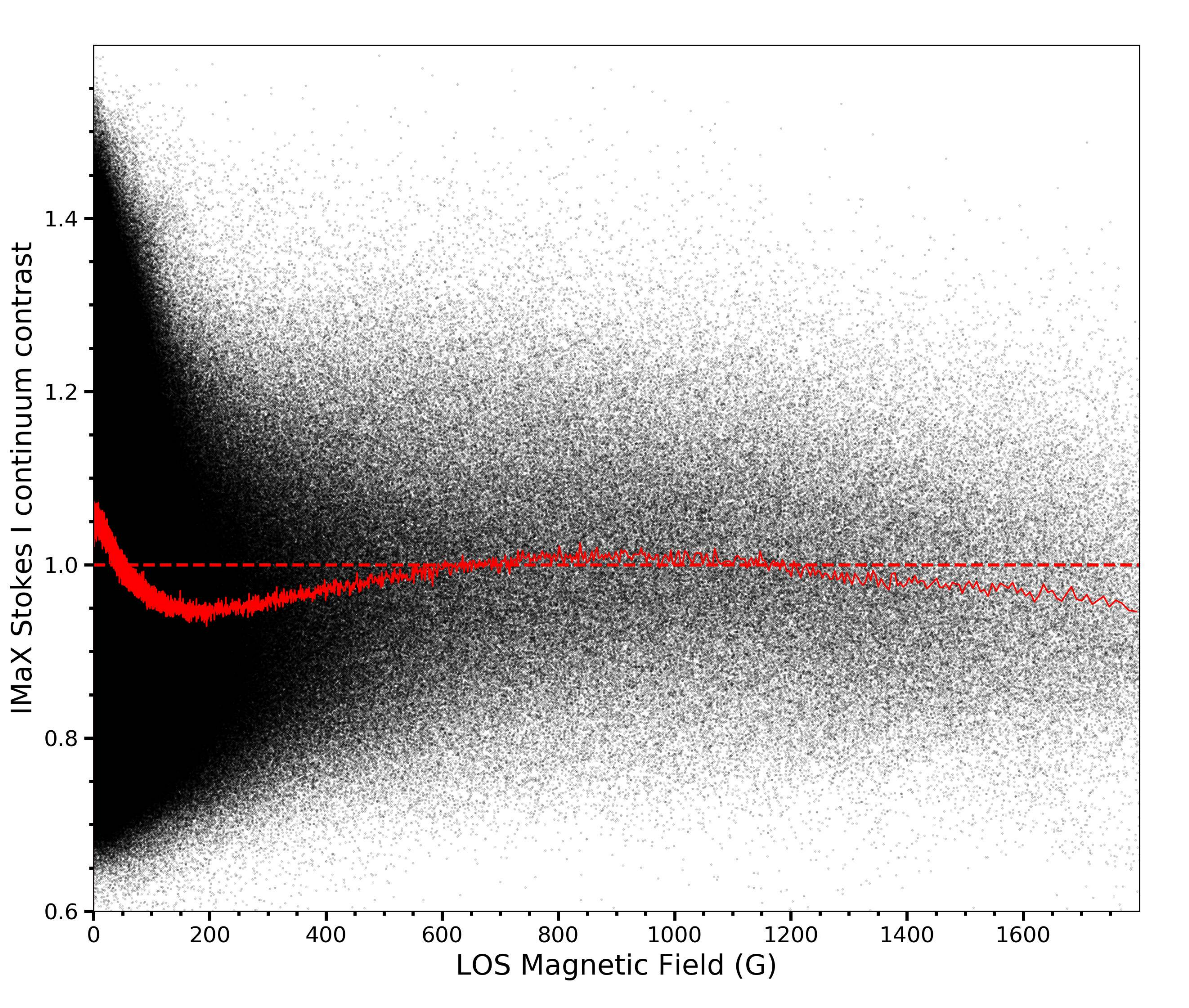}
\caption{Scatterplot of the continuum contrast  of Fe\,{\sc i} 5250.4\, \AA{} vs. the LOS component of the magnetic field, $B_{\rm LOS}$, for the AR plage. Pores and the FE region are excluded. The horizontal dashed red line is the mean QS continuum intensity level (the mean contrast of the pixels in the dashed red box in Figure~\ref{fig1}). The red curve joins the average contrast values in bins of 500 data points each.}
\label{imax_cont_sp}
\end{figure}

\begin{figure}
\centering
   \includegraphics[width=\hsize]{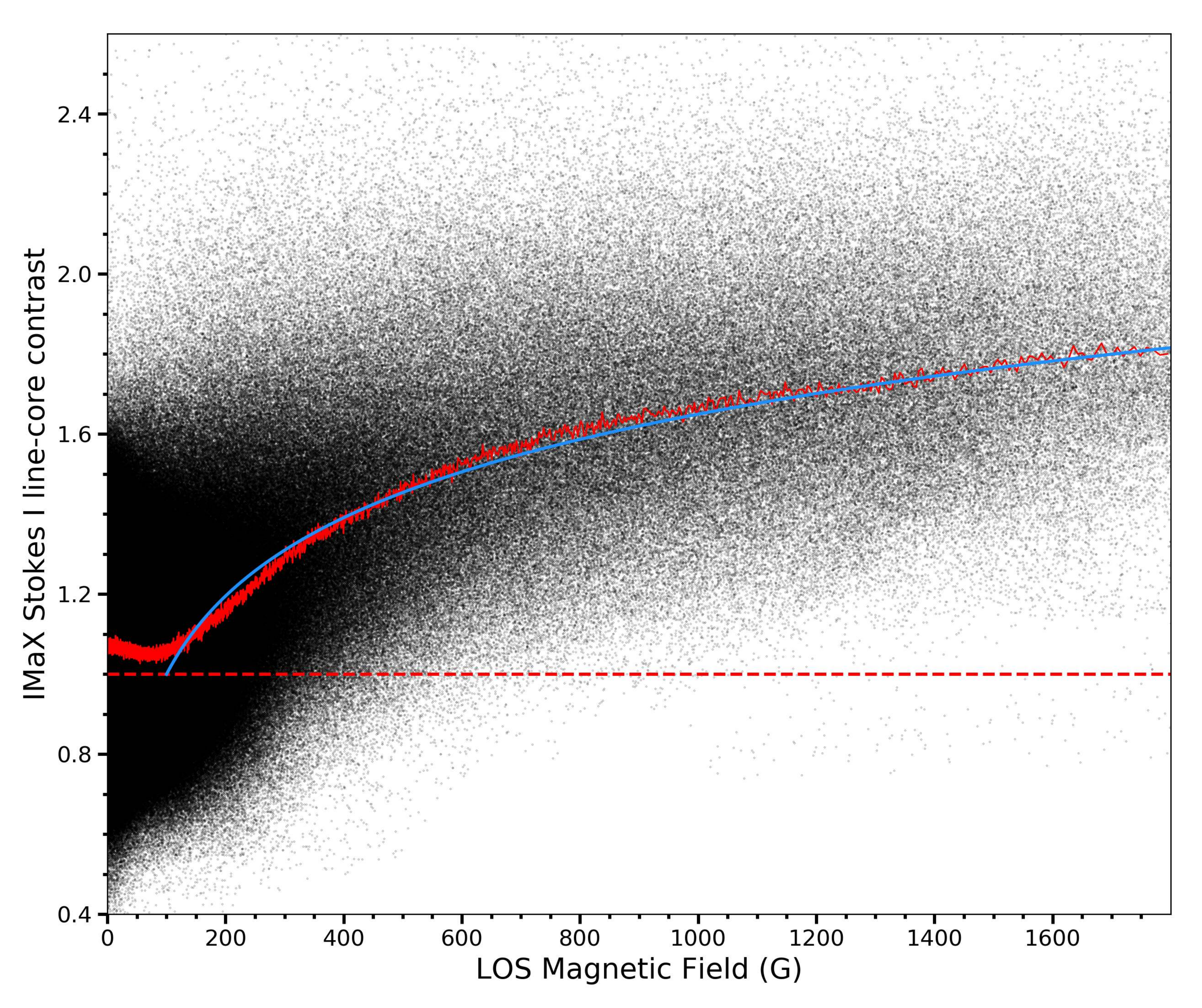}
\caption{Same as Figure~\ref{imax_cont_sp} but for the line core of Fe\,{\sc i} 5250.4\, \AA{}. The blue curve is the logarithmic fit to the averaged line-core contrast values starting at 120\,G.  }
\label{imax_lc_sp}
\end{figure}

After masking the pores, we plot in Figures~\ref{imax_cont_sp} and \ref{imax_lc_sp} pixel-by-pixel scatterplots of the contrast in the IMaX continuum and line-core of Fe\,{\sc i} 5250.2\, \AA{} versus the line-of-sight component of the magnetic field, $B_{\rm LOS}$. To show the trend of data points in the scatterplots, the contrast values are averaged into bins, each containing 500 data points. The averages for each bin are overplotted in red in both figures.\\

The large scatter around $B_{\rm LOS}=0$ in Figure~\ref{imax_cont_sp} is due to the distribution of the magnetic field in the granulation \citep{schnerr_brightness_2011}. At low magnetic field values ($B_{\rm LOS}$ < 150\,G), the contrast in the visible continuum decreases with increasing $B_{\rm LOS}$ since those fields are being dragged to the intergranular lanes by flux expulsion \citep{parker_kinematical_1963}. In an image, these pixels (not shown here) are located in the interiors of bright granules and dark intergranular lanes with weak magnetic fields. For fields above 200\,G, which correspond to magnetic features located in the intergranular lanes and undergoing convective instability, the contrast increases with $B_{\rm LOS}$ to reach contrast values above unity at 600\,G. The resulting shape of the curve at low fields is referred to as the fishhook \citep{schnerr_brightness_2011}.

Above 600\,G, the average continuum contrast is $>1$ and continues to increase until it reaches a maximum at $B_{max}=855$\,G, then decreases again, dropping below unity at around 1200\,G. The pixels belonging to this range (around the peak and after the turnover) are mainly located in the intergranular lanes, as is typical of small-scale magnetic elements \citep{solanki_small-scale_1993}. 
The numerical values of the longitudinal magnetic field obtained above (at the minimum of the fishhook and at the highest contrast, $B_{max}$) are computed by modelling the scattered data points using one of the non-parametric regression techniques described in the Appendix of Paper~I.
We prefer the non-parametric regression over a simple parametric model, such as the polynomial fit used in \citet{kobel_continuum_2011}, to find the contrast and $B_{\rm LOS}$ at the peak of their scatterplots, for two reasons. Firstly, the polynomial fit is highly dependent on the lower and upper limits of $B_{\rm LOS}$ within which the averaged contrast values are fitted. Secondly, it is also dependent on the degree of the fitting polynomial.

The IMaX continuum contrast in plage exhibits the same qualitative behaviour as the observations of \citet{frazier_multi-channel_1971}, \citet{narayan_small-scale_2010}, \citet{kobel_continuum_2011}, \citet{kostik_properties_2012}, \cite{Buehler_a_comparison} and the degraded simulations of \citet{danilovic_relation_2013}: a peak in the contrast at intermediate field strengths followed by a turnover at higher fields.  The main difference is that the magnetic field value at which the contrast reaches its maximum (850\,G) is higher than the values of 700\,G and 650\,G reported by \citet{kobel_continuum_2011} at their spatial resolution of 0.3" and by \citet{narayan_small-scale_2010} at 0.15", respectively. These values support the correlation between the position of the peak and the corresponding spatial resolution of the observations, as pointed out by \cite{yeo_intensity_2013}. 

The scatterplot of the line-core contrast versus $B_{\rm LOS}$ in Figure~\ref{imax_lc_sp} shows a weaker fishhook shape compared to the visible continuum. The monotonic increase of the line-core contrast with $B_{\rm LOS}$ was not seen in similar studies where line-core data were available \citep[]{title_differences_1992, yeo_intensity_2013}. Their scatterplots of the line-core contrast dropped at high magnetogram signals, even after masking out the pores.
Following paper I we model this relationship with a logarithmic function of the form $I(B) = \beta + \alpha.log B$, which is overplotted in blue in Figure~\ref{imax_lc_sp}. We cannot compare our fit results to any other study simply because the non-linear monotonic shape of the contrast in a similar wavelength band vs. $B_{\rm LOS}$ is not obtained by lower spatial resolution studies. Nevertheless, we perform them to serve as a reference to which future studies in the same wavelength band, or studies based on radiative MHD simulations, can be compared.
The IMaX continuum contrast could not be modeled with a logarithmic function given the non-monotonic increasing shape. This behaviour will be analysed in Section~\ref{turnover}.

\subsection{Scatterplots of UV contrast vs. $B_{\rm LOS}$}
\label{results_sufi}
\begin{figure}
\centering
\includegraphics[width=\hsize]{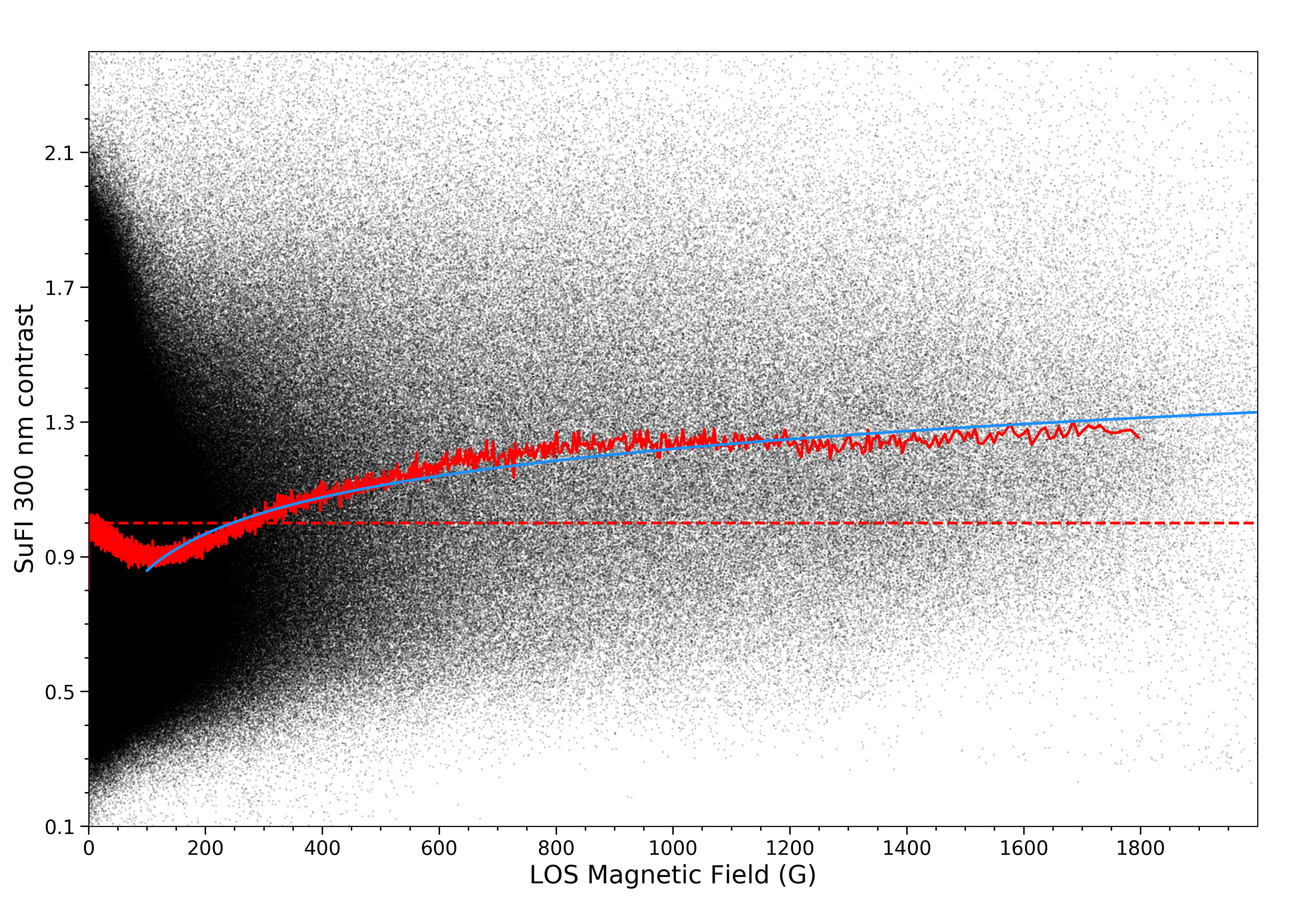}
\caption{Scatterplot of the SuFI intensity contrast at 300\,nm vs. $B_{\rm LOS}$. The red curve joins the averaged contrast values inside each bin, the blue curve is the logarithmic fit to the red curve starting from 100\,G.}
\label{sufi_300_sp}
\end{figure}

\begin{figure}
\centering
\includegraphics[width=\hsize]{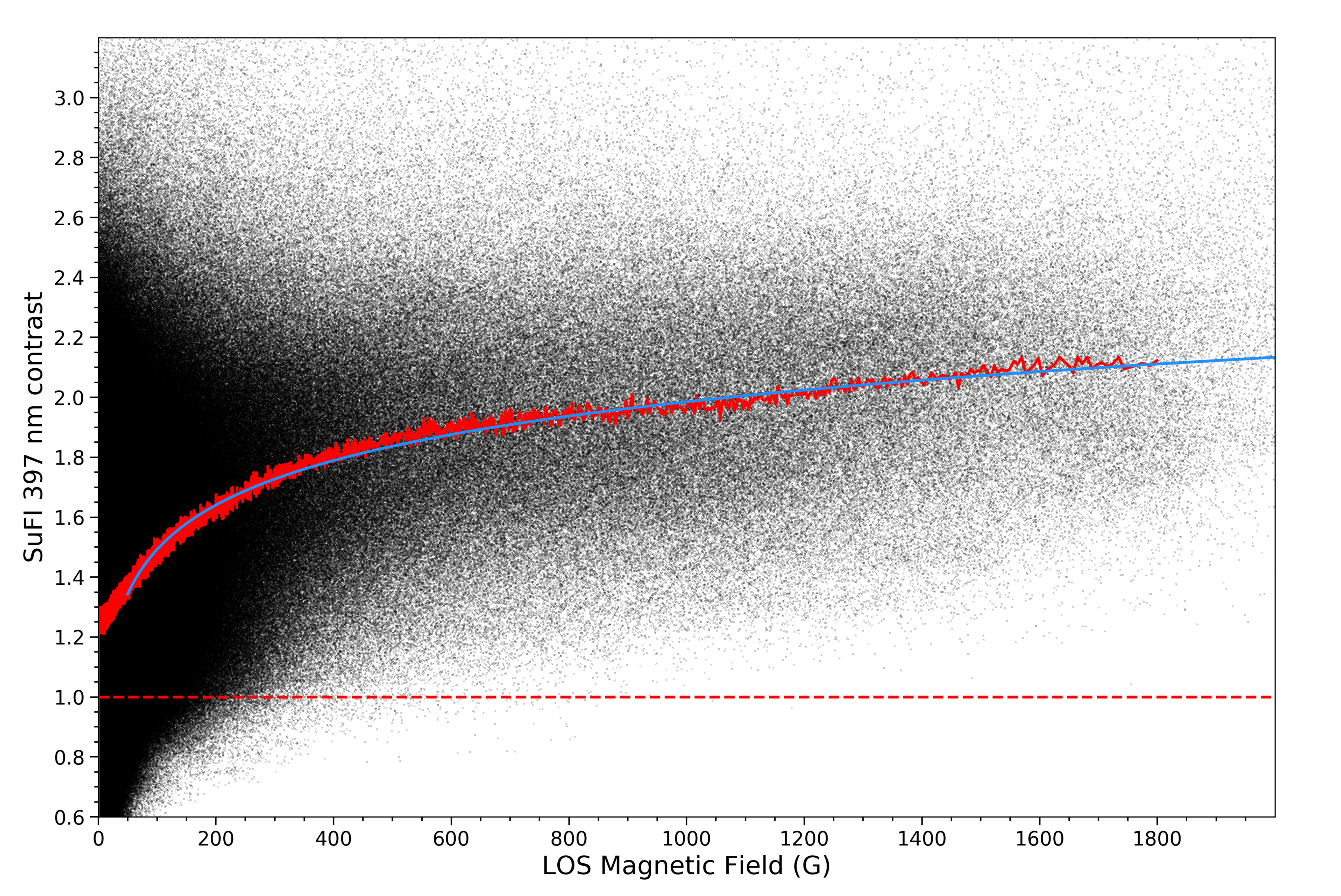}
\caption{Same as Figure~\ref{sufi_300_sp} but for the line core of Ca\,{\sc ii} H. The blue curve is the logarithmic fit to the red curve starting from 50\,G.}
\label{sufi_397_sp}
\end{figure}

Scatterplots of the contrast at 300 and 397\,nm against $B_{\rm LOS}$ are shown in Figures~\ref{sufi_300_sp} and \ref{sufi_397_sp}, respectively. Contrast values are averaged following the method described in Section~\ref{results_imax}.

At 300\,nm, the fishhook shape is still visible due to the relatively low formation height of this wavelength band of around 50\,km, as computed by  \cite{jafarzadeh_high-frequency_2017}. After an initial decrease with $B_{\rm LOS}$, the contrast at 300\,nm reaches its minimum  at around 100\,G. Beyond that point, the contrast increases with field strength until it saturates at higher $B_{\rm LOS}$ values.

The contribution function of the Ca\,{\sc ii} H bandpass, computed from an atmospheric model corresponding to AR plages and showing a mean formation height of 550\,km \citep{jafarzadeh_slender_2017}, implies that this line has a contribution from the photosphere, but also from the low-to-mid chromosphere.
The fishhook shape in the scatterplot of Figure~\ref{sufi_397_sp} is gone, due to the absence of granulation at the larger formation height of this line compared to the 300\,nm bandpass and the 525\,nm continuum. The features are on average brighter than the QS even at very weak fields, due to the presence of bright fibrils that are prominent even in weak magnetic field regions. Weak field regions that are dark can be seen mainly in the upper middle and lower left regions of the calcium images, while fibrils are found nearly everywhere (see Fig.~\ref{fig2}). The contrast then increases with $B_{\rm LOS}$ until it saturates at higher fields. Pixels in this region of the scatterplot belong to plages as identified in the IMaX continuum images. 

In the past, a power-law function has been used to describe the relation between Ca\,{\sc ii} H brightness and magnetic flux density for network and IN features both, within limited fields of view \cite[]
{rezaei_relation_2007,loukitcheva_relationship_2009}  
and for magnetic features in full-disk observations \citep[]{ortiz_how_2005, chatzistergos_analysis_2017}. Here, however, we follow Paper~I, where we found that a logarithmic function provides a superior representation to the data at the high spatial resolution reached by $\sunrise$. The fit parameters are shown in Table~\ref{tab1}.

\begin{table}
\caption{Parameters of logarithmic fits to the contrast in the core of Fe\,{\sc i} line at 5250.2\, \AA{} (starting from 120\,G) and for 300 nm and 397 nm (starting from 100\,G) vs. $B_{\rm LOS}$. }
\label{tab1}
\centering
\begin{tabular}{c c c c}
\hline \hline

Wavelength band & $\alpha$ & $\beta$ & $\chi^2$ \\
\hline
IMaX line core & 0.673$\pm$ 0.002 & -0.365$\pm$ 0.004 & 9.232 \\
\hline
300 & 0.353 $\pm$ 0.002&0.156$\pm$ 0.005&9.213\\
\hline
397 & 0.512$\pm$ 0.001&0.456$\pm$ 0.003&5.55\\
\hline
\hline
\end{tabular}

\end{table}

\subsection{Comparison between wavelengths}
\label{all_wav}

\begin{figure*}[h!]
\centering
\includegraphics[width=\hsize]{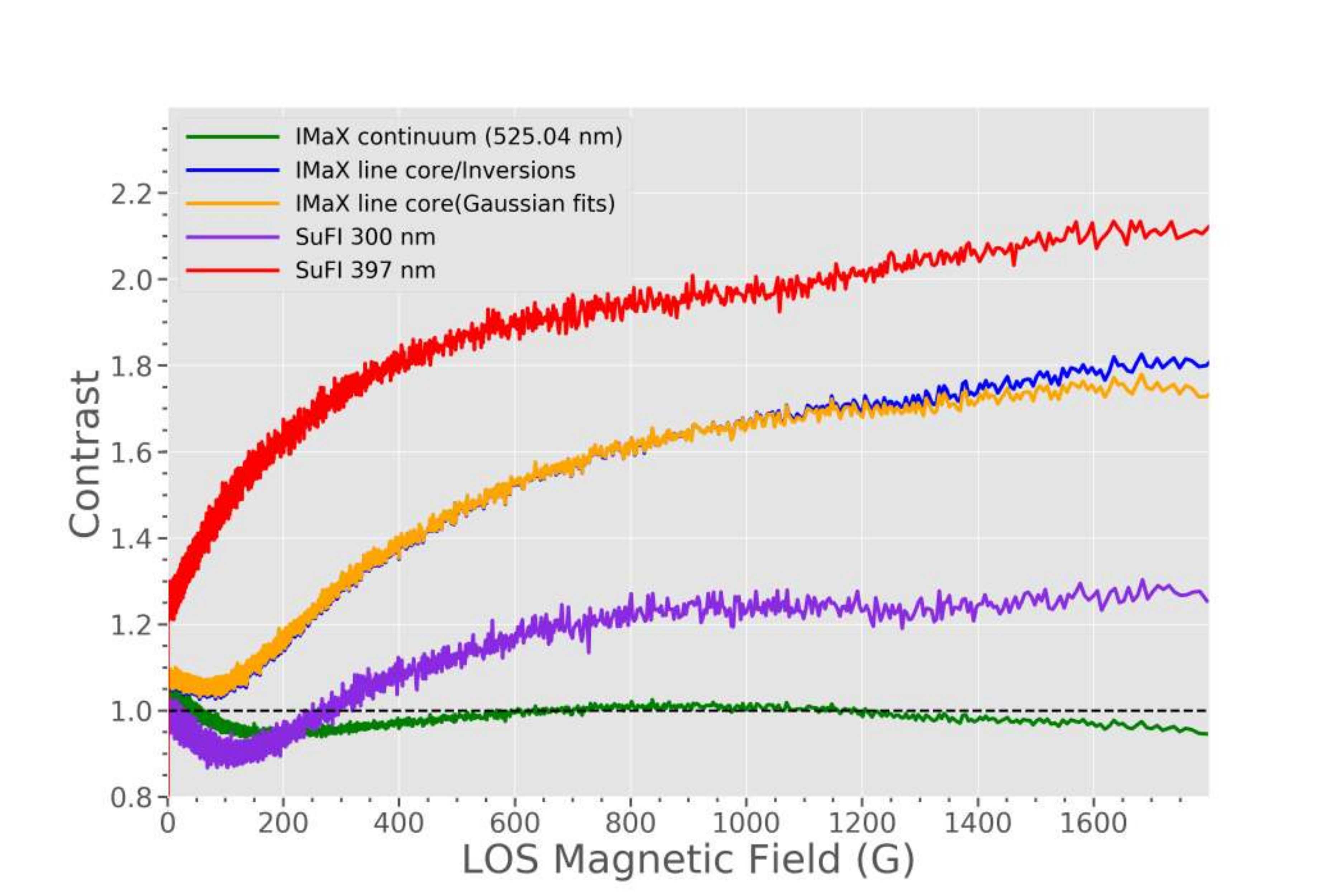}
\caption{Intensity contrasts of plage vs. $B_{\rm LOS}$ in the wavelengths considered in the current study (see the legend in the figure). Plotted are the contrast values (plotted in red in Figures~\ref{imax_cont_sp}, \ref{imax_lc_sp}, \ref{sufi_300_sp} and \ref{sufi_397_sp}). For comparison, also plotted is the binned line-core contrast at 525.02\,nm derived from Gaussian fits to individual Stokes~$I$ profiles.}
\label{all}
\end{figure*}

\begin{figure*}[ht!]
\centering
\includegraphics[width=\textwidth]{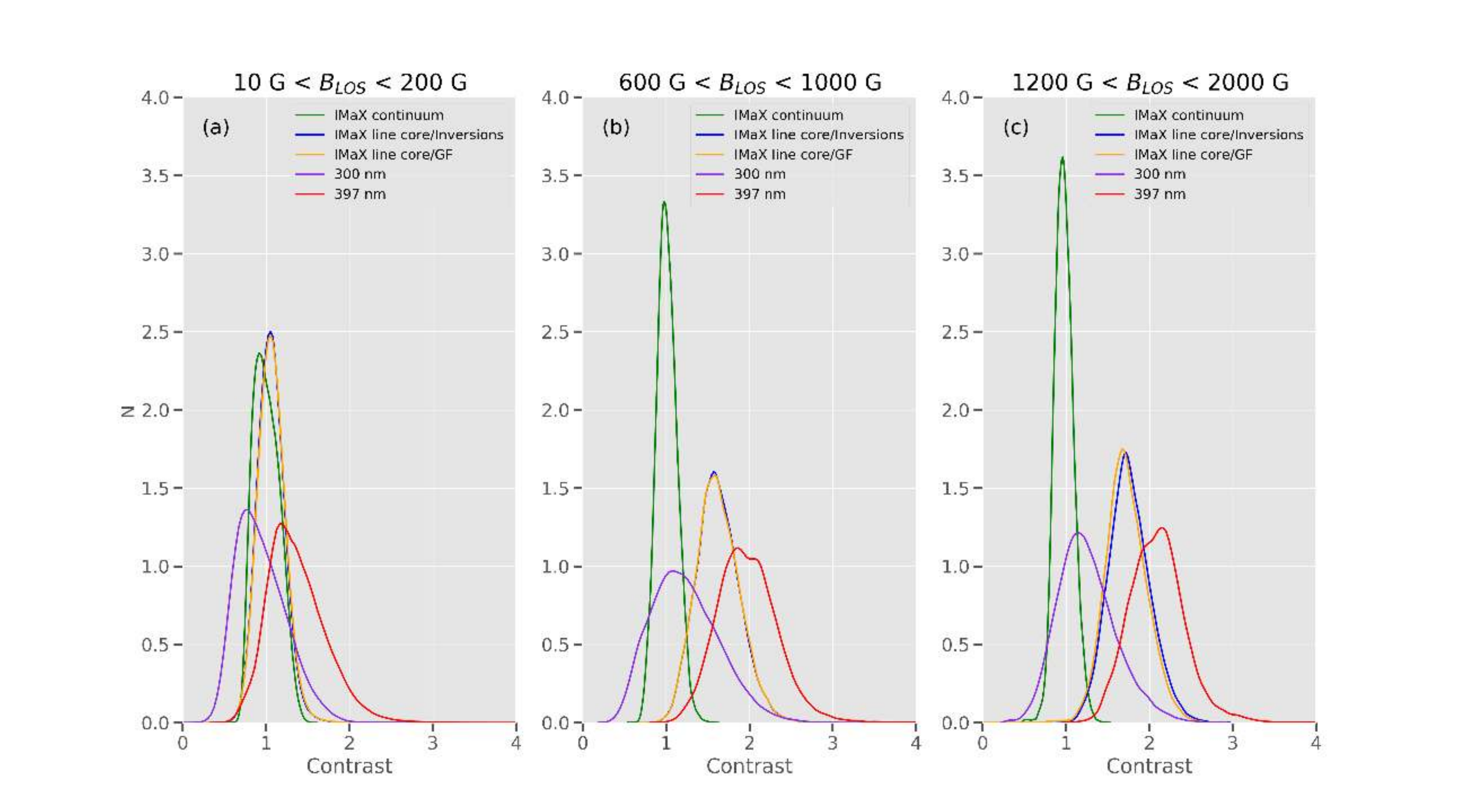}
\caption{Contrast histograms of pixels belonging to the common FOV of SuFI and IMaX after masking out pores. Colors correspond to different wavelengths, and panels correspond to different ranges of $B_{\rm LOS}$ values (see text for a detailed explanation). The histograms are normalized such that the integral over each is equal to one.  }
\label{hist}
\end{figure*}

We plot in Figure~\ref{all} the binned contrast curves for the wavelengths in the UV at 300 and 397\,nm, and in the visible continuum at 525.04\,nm plus line core of Fe\,{\sc i} at 525.02\,nm derived from the inversions as explained in Section~\ref{contrast}. We plot for comparison the curve corresponding to the line-core contrasts derived from the Gaussian fits to individual IMaX Stokes~$I$ profiles. 
Histograms of the measured contrasts at the studied wavelengths are displayed in Figure~\ref{hist}. These histograms are for the common FOV ($13^{\prime\prime}\times34^{\prime\prime}$) of IMaX and SuFI data sets for a one to one correspondence of the contrasts. All the pixels of the time series except for those belonging to pores are included in these histograms. They are plotted for different ranges of $B_{\rm LOS}$, these ranges correspond to magnetic field values where the scatterplot of the contrast at 525.04 nm vs. $B_{\rm LOS}$ shows different behaviour: the weak field region (10\,G < $B_{\rm LOS}$ < 200\,G), the range where features have an average contrast above unity (600 G < $B_{\rm LOS}$ < 1000\,G), and the range where features start to be darker than the mean quiet-Sun intensity ($B_{\rm LOS}$ > 1200\,G).

For weak fields (Fig.~\ref{hist}a), 525\,nm continuum (green) and 300 nm (violet) have averaged contrasts < 1. Both bands exhibit lower contrasts than the rest of the wavelengths at all field strengths, with the contrast at 300 nm becoming larger than 525 nm with increasing $B_{\rm LOS}$ (Figs.~\ref{hist}b and c).

The higher contrast measured in the UV (at 300 and 397 nm) compared to the visible continuum at 525.04\,nm is explained by the stronger response of the Planck function to the temperature variations at shorter wavelengths plus the greater height of formation in the case of the Ca\,{\sc ii} H line core. For the Fe\,{\sc i} line core, the reason for the larger contrast, is this line's great temperature sensitivity and its greater formation height. For all values of $B_{\rm LOS}$, Ca\,{\sc ii} H (red) exhibits the highest contrasts.

Figures~\ref{all} and \ref{hist} also show the difference in the computed line-core intensity from the inversions (blue) and the Gaussian fits (orange), which is more notable at higher fields (Fig.~\ref{hist}c). As explained in Section~\ref{contrast}, this difference is due to the underestimation of the line-core intensity at higher fields if derived from the Gaussian fits. The curves obtained by the two methods agree rather well, except at the very highest field strengths, suggesting that the Zeeman splitting plays a minor role at the spectral resolution of IMaX.
 \subsection{Why does the continuum contrast display a turnover?}
\label{turnover}

\begin{figure*}[ht!]
\centering
\includegraphics[scale=.65]{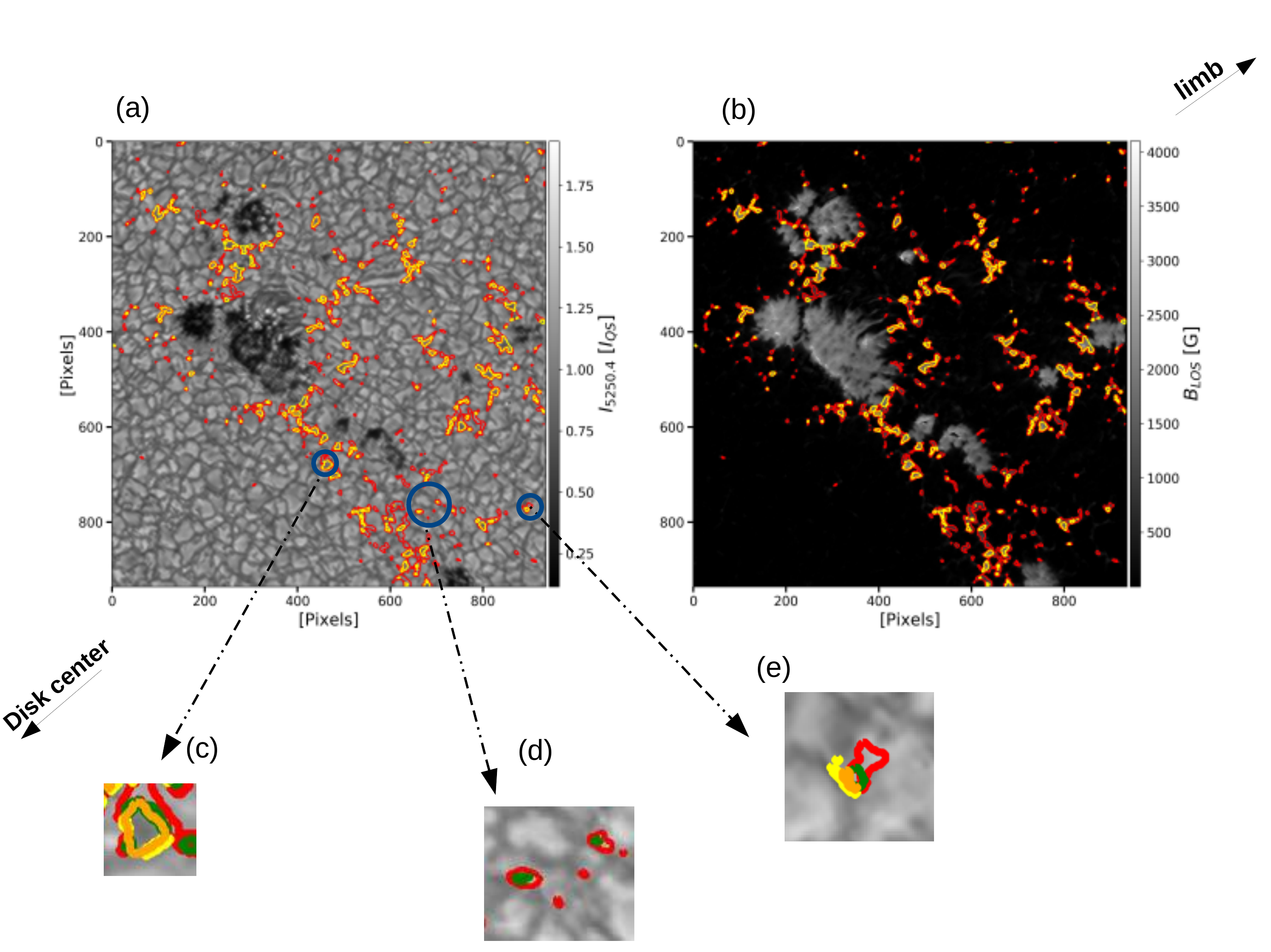}
\caption{(a), (b): IMaX data for the same time as in Figure~\ref{fig1}, now with the absolute value of $B_{\rm LOS}$ plotted in the right panel. The red contours enclose the `peak' pixels with $>1$ averaged contrast in the continuum of 525 nm and 600\,G < $B_{\rm LOS}$ < 1000\,G, while the yellow contours enclose the `turnover' pixels with averaged contrasts $<1$ and $B_{\rm LOS}$ larger than 1200\,G. (c), (d), (e): Blowups of three magnetic features of different sizes. Red and green contours enclose the bright (contrast > 1) parts of both families of pixels, while orange and yellow contours enclose the dark (contrast < 1) parts.
}
\label{contour}
\end{figure*}
In this subsection we will try to understand why the averaged continuum contrast of plage features at 525\,nm peaks at intermediate field strengths and decreases at higher fields when observed by IMaX near disk center. 
\subsubsection{Treatment of magnetic signals around pores} 
The derived contrast as a function of $B_{\rm LOS}$ near disk center is believed to vary according to the treatment of pixels around pores and sunspots \citep[]{kobel_continuum_2011,yeo_intensity_2013}. Therefore, care has to be taken when removing pores since plages are often located close to their boundaries, and inclusion of pixels which might belong to pores could affect the contrast-$B_{\rm LOS}$ relationship in the plage areas.
We test this by extending our masks around the pores, described in Section~\ref{masking}, to exclude more of their surroundings. We then again produce the scatterplot to see if the shape described above persists or not. The shape of the scatterplot (not shown here) and in particular the position of the contrast peak were hardly affected. 

\subsubsection{Masking technique}
The criterion we use here to exclude pores from our data is based solely on the IMaX continuum intensity (see Sec.~\ref{masking}).
We also try a mask based on a contrast-$B_{\rm LOS}$ threshold, such as that employed by \cite{kobel_continuum_2011}. The threshold used for creating their masks (contrast < 0.85 and $B_{\rm LOS}$ > 900\,G) aimed to remove the inner dark and magnetic parts of pores. In addition, they extended their masks around pores to exclude the pixels darkened by telescope diffraction. However, in our data this threshold corresponds to micropores that are part of the plage, and also to the dark magnetic pixels found in the ribbon-like features which result from their possible inclination with respect to the solar normal (see Sec.~\ref{morphology}).
Thus, adopting the same contrast-magnetic field threshold in our data will remove the pixels contributing to the downturn in the bottom right part of our scatterplot in Figure~\ref{imax_cont_sp} in addition to a few brighter pixels if those masks are extended, since micropores in our data tend to have brighter edges (see Figs.~\ref{profiles}a and b). This points to a dependence of the photospheric continuum contrast-$B_{\rm LOS}$ relationship on the criterion used to distinguish between magnetic elements and pores. This also implies that a careful examination of the magnetic/photometric distribution of small-scale magnetic elements has to be done before adopting any masking criterion, which is only possible if features are well resolved, as is the case with IMaX (see Sects.~\ref{morphology} and \ref{QS_2009_2013}).  

\subsubsection{Spatial resolution}
\label{morphology}
In lower spatial resolution studies \citep[]{title_differences_1992,topka_properties_1992,lawrence_contrast_1993} using the SVST, the continuum contrast monotonically decreased with magnetogram signal (see Introduction). The inability to resolve magnetic regions, or the fact that they are partly blended with the non-magnetic surroundings, primarily with the dark intergranular lanes, were considered the reason for the negative contrasts at higher field strengths. 
This was confirmed by \cite{danilovic_relation_2013} using MHD snapshots of a plage region. After degrading the simulations to the spatial resolution of Hinode/SP, their scatterplots turned from a monotonic relationship at the original MuRAM resolution to a peak and a turnover, in agreement with the findings of \cite{kobel_continuum_2011}.

\cite{danilovic_relation_2013} showed that smearing isolated bright magnetic structures embedded in a dark, nearly field-free environment shifts the data points with high contrast and $B_{\rm LOS}$ at the original resolution to a region of the scatterplot with lower contrasts and intermediate field strengths. The accumulation of data points there gives rise to a peak upon averaging the contrast values. Bigger and less bright magnetic structures are only slightly affected by the smearing and remain dark with higher field strengths, causing the turnover in the right side of the scatterplot. 
This interpretation of the effect of finite spatial resolution is valid for regions with low average magnetic field such as the quiet-Sun network, where the magnetic features have little internal structure. Thus, it works well for the data described in Paper~I. However, The plage features in our data are larger than those found in the quiet Sun, so that they should be affected less by spatial smearing, as test have confirmed. Instead the magnetic features in the plage show more complex photometric and magnetic distributions than the simple picture of magnetic bright points surrounded by dark field-free lanes. 

To isolate these structures, we locate the pixels with $B_{\rm LOS}$ > 1200\,G and binned contrast < 1 as seen in Figure~\ref{imax_cont_sp}. We call them the `turnover' pixels. We also locate the pixels centered at $\pm$ 200\,G from $B_{\rm max}$ (i.e., 600\,G < $B_{\rm LOS}$ < 1000\,G) with an averaged contrast > 1, and call them the `peak' pixels. Figures~\ref{contour}a and b show the contours enclosing both families of pixels. We find that together they form magnetic structures of different sizes and appearences embedded in intergranular lanes and spread mainly around the large and small pores, with the turnover pixels (yellow contours) typically being surrounded by the peak pixels (red contours). Small structures have the shape of circular bright points, while slightly larger ones resemble the so called `striation' or `ribbon' in the G-band observations of \cite{berger_solar_2004}. Figs.~\ref{contour}c, d, and e are blowups of three features, picked to illustrate three distinctive types: a micropore (Fig.~\ref{contour}c), bright points (Fig.~\ref{contour}d), and a ribbon-like feature (Fig.~\ref{contour}e). The red and yellow contours now enclose the bright (contrast > 1) and dark (contrast < 1) parts of the peak pixels, while the green and orange contours enclose the bright and dark parts of the turnover pixels. Clearly, the smallest features, the bright points in Fig.~\ref{contour}d, contain only bright pixels, while the larger features display an increasing fraction of dark pixels as their area increases. This behaviour is found to be typical. 

\begin{figure*}[ht!]
\centering
\includegraphics[width=\textwidth]{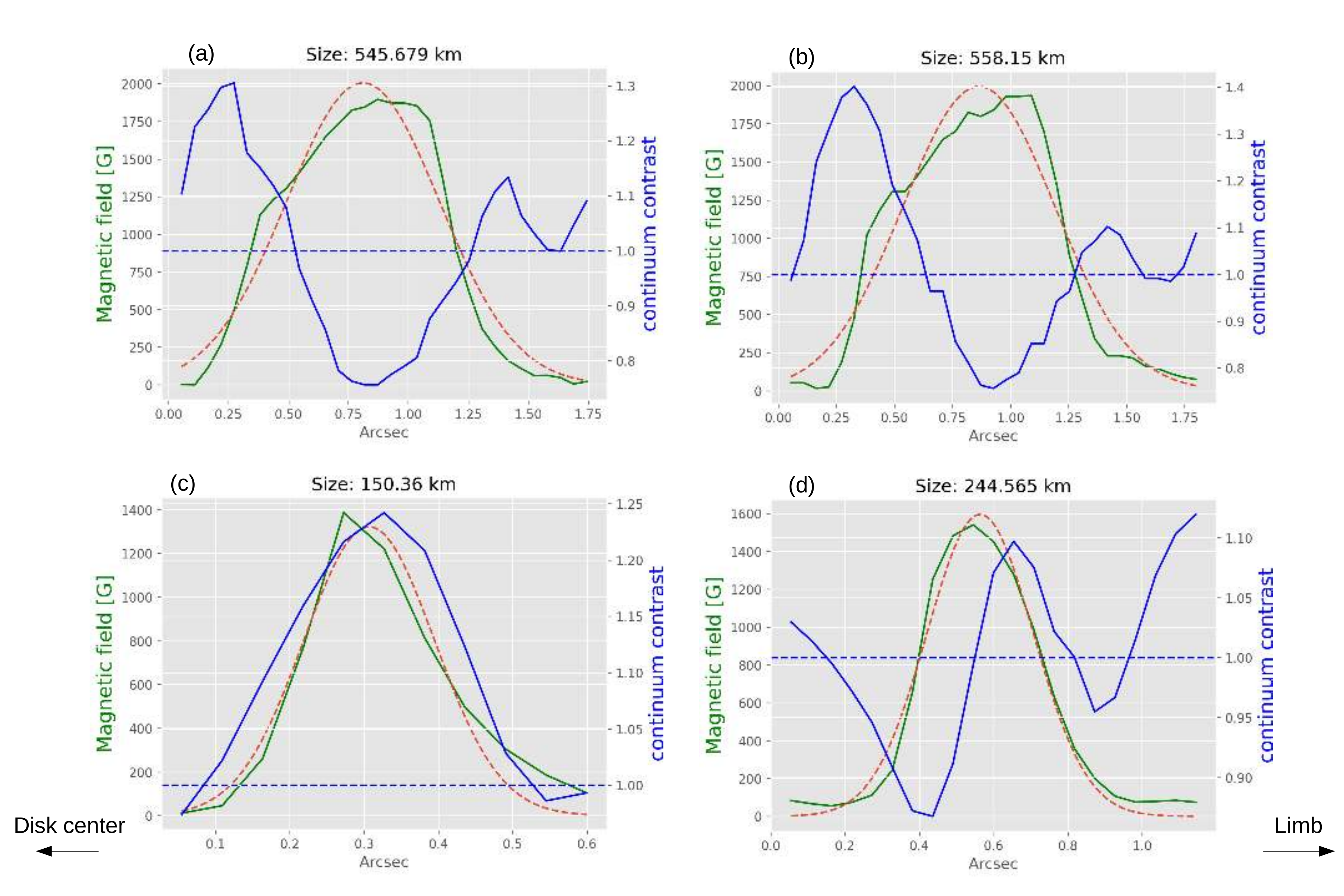}
\caption{Profiles of $B_{\rm LOS}$ (green curve) and continuum intensity contrast at 525\,nm (blue curve) along cuts through 4 different magnetic structures in an example IMaX image (see text for details). The red dashed curves are the Gaussian fits to the $B_{\rm LOS}$ profiles. The FWHM of these Gaussian fits are given at the top of each frame. The horizontal dashed blue line is where the contrast is equal to unity. The limb is to the right of the figure, while disk center is to the left.}
\label{profiles}
\end{figure*}

In Figure~\ref{profiles} we plot horizontal cuts through 4 features belonging to the three types of magnetic elements shown in Figure~\ref{contour}. Displayed are the continuum intensity at 525 nm and the LOS magnetic field. The cut is perpendicular to the limb, with the nearest limb being to the right of the figure, and disk center to the left.

In all 4 features the LOS component of the magnetic field peaks in their centers and decreases towards their edges. The continuum contrast at 525\,nm, however, shows three distinctive patterns according to the sizes of the features (given above each frame): for large features (horizontal extent of 500\,km-600\,km) it is > 1 at the edges and < 1 at the center where $B_{\rm LOS}$ peaks (Figs.~\ref{profiles}a and b). For small bright points (100\,km-200\,km), it is > 1 in their centers and the brightness peak coincides with the position where $B_{\rm LOS}$ peaks (Fig.~\ref{profiles}c). In the elongated features with intermediate sizes (200\,km-400\,km), the continuum contrast is > 1 in the limb side of the feature, and < 1 in the disk center side, with a neutral contrast in the center (Fig.~\ref{profiles}d). The latter pattern is seen in most of the features in the  IMaX data.
The feature size is determined from a Gaussian fit (dashed red curves in Fig.~\ref{profiles}) to the $B_{\rm LOS}$ profile.

The profiles of the continuum intensity in Figs.~\ref{profiles}a and b resemble the symmetric double-humped profiles at disk center in both G-band observations and numerical simulations \citep{berger_solar_2004, steiner_recent_2005}. While the profile in Fig.~\ref{profiles}d resembles the asymmetric pattern of contrast in magnetic elements near the limb \citep{hirzberger_solar_2005,steiner_recent_2005}. Given that our observations are not carried out exactly at disk center ($\mu$ = 0.93), it could be that the shape of the intensity profiles seen across the mid-sized features are due to an inclined line of sight (the hot wall effect). But it could also be a result of inclined magnetic fields with respect to the solar normal, as proposed in \cite{keller_origin_2004}.

Variations of the intensity contrast and magnetic field distributions across IMaX features with different sizes explain the scatter we see in the plot of IMaX continuum contrast vs. $B_{\rm LOS}$ in the range 600\,G-2000\,G, which corresponds to magnetic elements belonging to plage. We conclude that the scatter around the mean QS level is real and is not a consequence of poor spatial resolution or noise in the data. This point is strengthened later when inspecting the brightness-magnetic field relationship in quiet-Sun areas located within the FOV of the AR observations (Section~\ref{QS_2009_2013}).

\subsection{Comparison with 2009 quiet-Sun data}
\label{QS_2009_AR}

In this section we present a qualitative and quantitative comparison of the contrast-magnetic field relationship near disk center in AR plage with that of quiet Sun, as obtained in Paper~I. This comparison is obviously restricted to the wavelengths that were recorded during both flights (IMaX continuum and line core, 300\,nm continuum, and Ca\,{\sc ii} H line core at 397\,nm). The reason we don't compare with the quiet regions extracted from the active-regions scans is that the quiet Sun is located mainly outside the SuFI FOV. In the next section we will analyse the IMaX properties of quiet-Sun features from 2013 scans, and compare them to the 2009 results to test for consistency. 

\begin{figure*}[ht!]
\centering
\includegraphics[width=\textwidth]{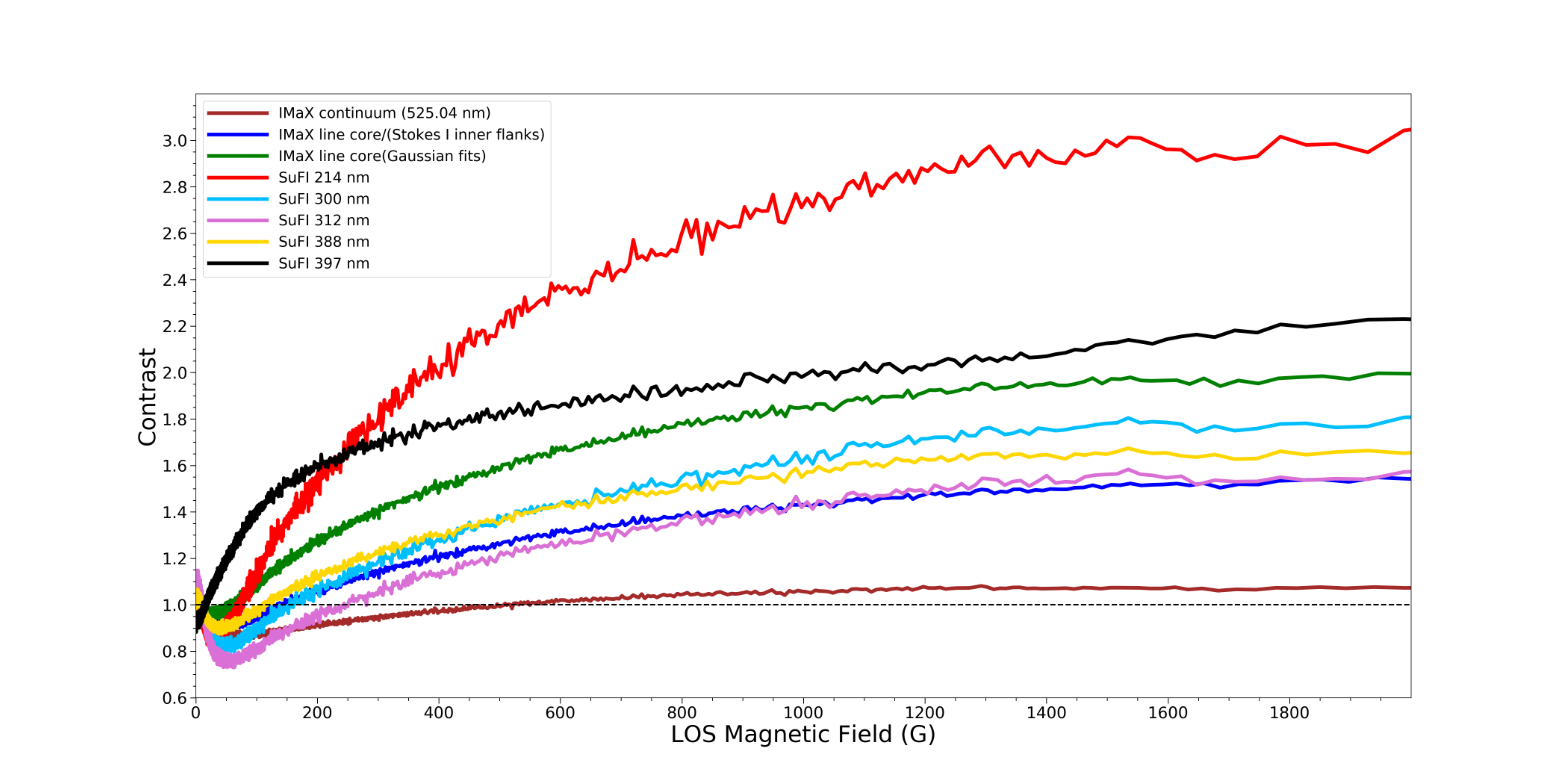}
\caption{All binned contrast vs.~$B_{\rm LOS}$ curves from Figures 3, 4, 7 and 8 of Paper~I plotted together. Also plotted is the contrast of the 5250\,\AA{} line core obtained by averaging the intensities at the wavelength positions +40 and -40 m\,\AA{} from line core. The curves are identified by their color in the legend in the upper left part of the figure. The black dashed line marks the mean quiet-Sun intensity level, i.e., a contrast of unity.}
\label{bins_qs_updated}
\end{figure*}

We show in Figure~\ref{bins_qs_updated} the averaged contrasts vs. $B_{\rm LOS}$ obtained from the quiet-Sun observations in 2009 (updated Figure~9 of Paper~I). Some of these curves differ from those plotted in Paper I, due to an error in the SuFI straylight removal in that paper.

The main differences to the plot in Paper~I are listed in the points below:
\begin{itemize}
\item The QS Ca\,{\sc ii} H (black) reaches higher contrast than the line core of IMaX (green). This is basically due to the fact that the Ca\,{\sc ii} H contrast (SuFI observations) increased relative to Paper~I, while the IMaX line-core contrast remained unchanged. 
\item The contrast at 313\,nm (pink) is still lower than 388\,nm (yellow) but the difference is less pronounced than obtained in Paper~I.
\item The contrast at 300\,nm (light blue) is now higher than 388\,nm (yellow) for $B_{\rm LOS}$ above 600\,G.
\end{itemize}

\begin{figure*}[ht!]

\centering
\includegraphics[width=\textwidth]{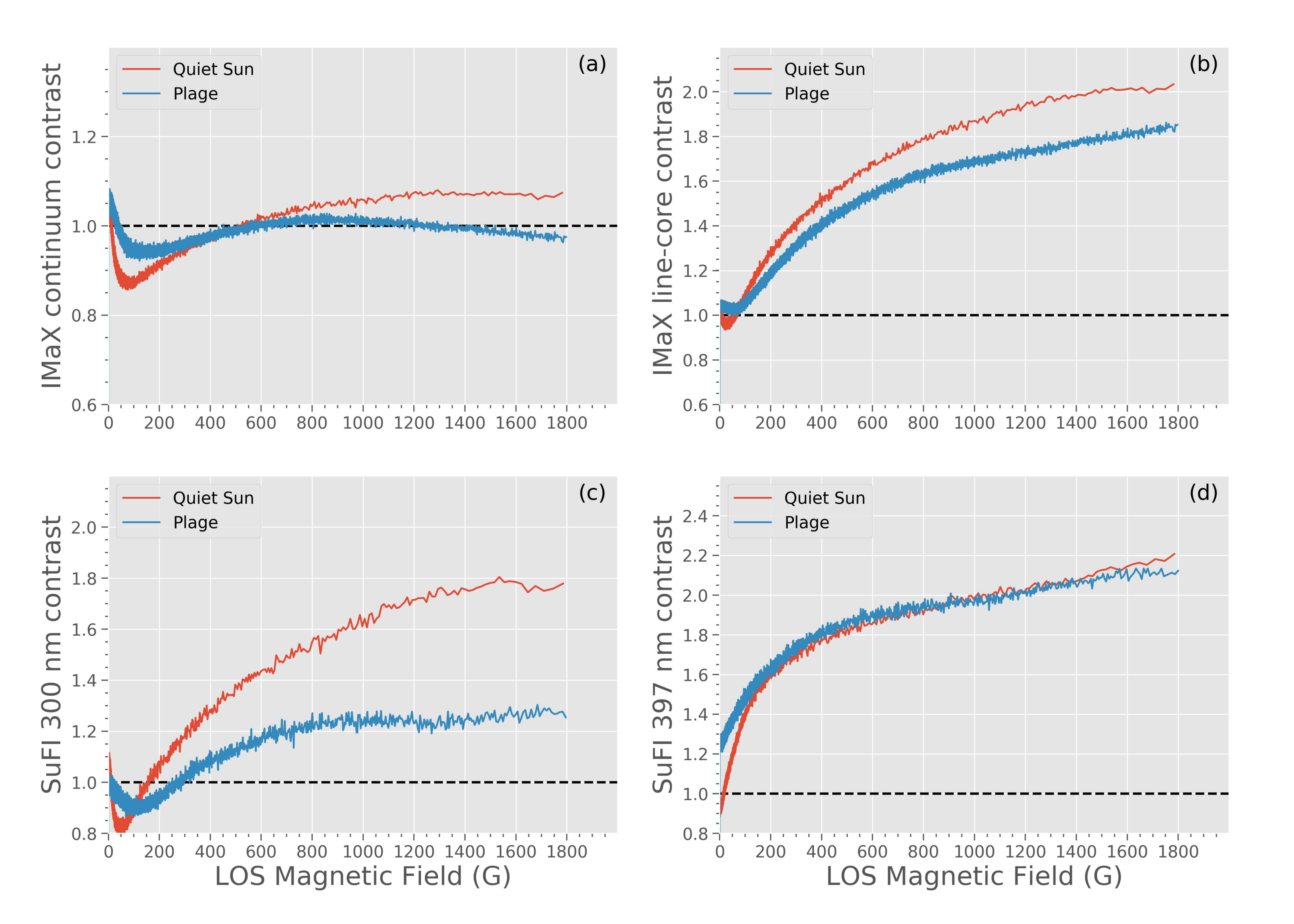}
\caption{Averaged contrast of (a) IMaX continuum, (b) IMaX line core, (c) SuFI 300 nm, (d) SuFI 397 nm against $B_{\rm LOS}$ for the quiet Sun observed by $\sunrise$ in 2009 (blue curves) and for plage (red curves) in 2013.}
\label{qs_plage}
\end{figure*}

In Figure~\ref{qs_plage}, we overplot the averaged contrasts for the quiet Sun (red curves) observed with $\sunrise$~I and analysed in Paper~I (updated and consistent with the curves shown in Figure~\ref{bins_qs_updated}) and for the AR plage (blue curves) analysed in the current work, of the IMaX continuum (panel (a)) and line core (panel (b)) at 525 nm, the continuum of 300 nm (panel (c)), and the line core of Ca\,{\sc ii} H at 397 nm (panel (d)) against the photospheric magnetic field. The black dashed line corresponds to a contrast of 1.

In all 4 panels, the plage curves lie generally higher than the QS ones at low $B_{\rm LOS}$, but are lower at larger $B_{\rm LOS}$, with the latter effect being marginal for Ca\,{\sc ii} H. Very close to $B_{\rm LOS}=0$, however, the QS curves are higher again for SuFI 300\,nm and for IMaX continuum.
For SuFI 300\,nm and IMaX continuum the larger variation of the contrast curve in QS at small $B_{\rm LOS}$ is likely due to the brighter granules and darker intergranular lanes than in the abnormal granulation present in the plage region \citep{narayan_small-scale_2010}. This `fishhook shape' \citep{schnerr_brightness_2011} is extended to larger $B_{\rm LOS}$ in plage due to the filling of intergranular lanes (and to some extent the sides of granules) with weak field. The greater brightness of the IMaX continuum in the QS is in accordance with the findings of \cite{title_differences_1992}, \cite{lawrence_contrast_1993} and \cite{kobel_continuum_2011}. Whereas the transition to higher QS contrast in magnetic features occurs at 500\,G for the IMaX continuum, it occurs already at around 100\,G for the IMaX line core and the SuFI 300\,nm channel. In addition, plage magnetic elements at 300\,nm reach a contrast >1 at around 250\,G, while QS magnetic elements make this transition already at 150\,G. The brightness in the Ca\,{\sc ii} H line core in both, quiet-Sun network and plage is similar for $B_{\rm LOS}$ above 1000\,G. The slightly higher QS contrast above approximately 1400\,G is based on too few QS points to be truly significant.
 
In the literature, the higher contrasts at larger $B_{\rm LOS}$ values reached by the visible continuum in the QS than in AR plages was attributed to the higher efficiency of convective energy transport in the quiet Sun. Going from QS to AR, i.e. with increasing magnetic flux averaged over the region, convection is suppressed in the surroundings of the stronger and larger magnetic features (ARs) leading to less efficient heating from the convective walls, and therefore less brightening with respect to their counterparts in the quiet Sun \citep[]{vogler_effect_2005,morinaga_suppression_2008,kobel_continuum_2012,criscuoli_comparison_2013,riethmuller_comparison_2014}.
The same reasoning may also explain the significantly lower contrast in plage at larger $B_{\rm LOS}$ seen at 300\,nm, given the similar formation height to the IMaX continuum.

In the cores of spectral lines (Figs.~\ref{qs_plage}b and d), this difference is less significant compared to features formed in lower layers. This result is in accordance with the temperature models of network and plage flux tubes derived from the FTS observations by \cite{solanki_solar_1996}, cf.~\cite{solanki_photospheric_1987} and \cite{solanki_continuum_1992}. In these works, flux tubes in the network were found to be hotter than in plage, especially in lower regions of the photosphere.

\subsection{Comparison between 2009 and 2013 quiet-Sun data}
\label{QS_2009_2013}

As mentioned in Section~\ref{inversions}, we can identify some quiet-Sun areas in IMaX scans (see for example red box in Figure~\ref{fig1}). We also delimit a couple of quiet-Sun areas in the upper right side of the FOV. We produce contrast-$B_{\rm LOS}$ scatterplots of the pixels embedded in these boxes to test if we can reproduce the relationship obtained from observations of the solar surface when the Sun was mostly quiet, i.e $\sunrise$ observations from 2009. For brevity, we refer to QS data from 2009 as QS-2009 and QS boxes from scans recorded in 2013 as QS-2013. Since the quiet-Sun boxes in 2013 data are outside the SuFI field of view, we restrict this comparison to the IMaX continuum and line-core contrasts vs. $B_{\rm LOS}$.

\begin{figure*}[ht!]
\centering
\includegraphics[width=\textwidth]{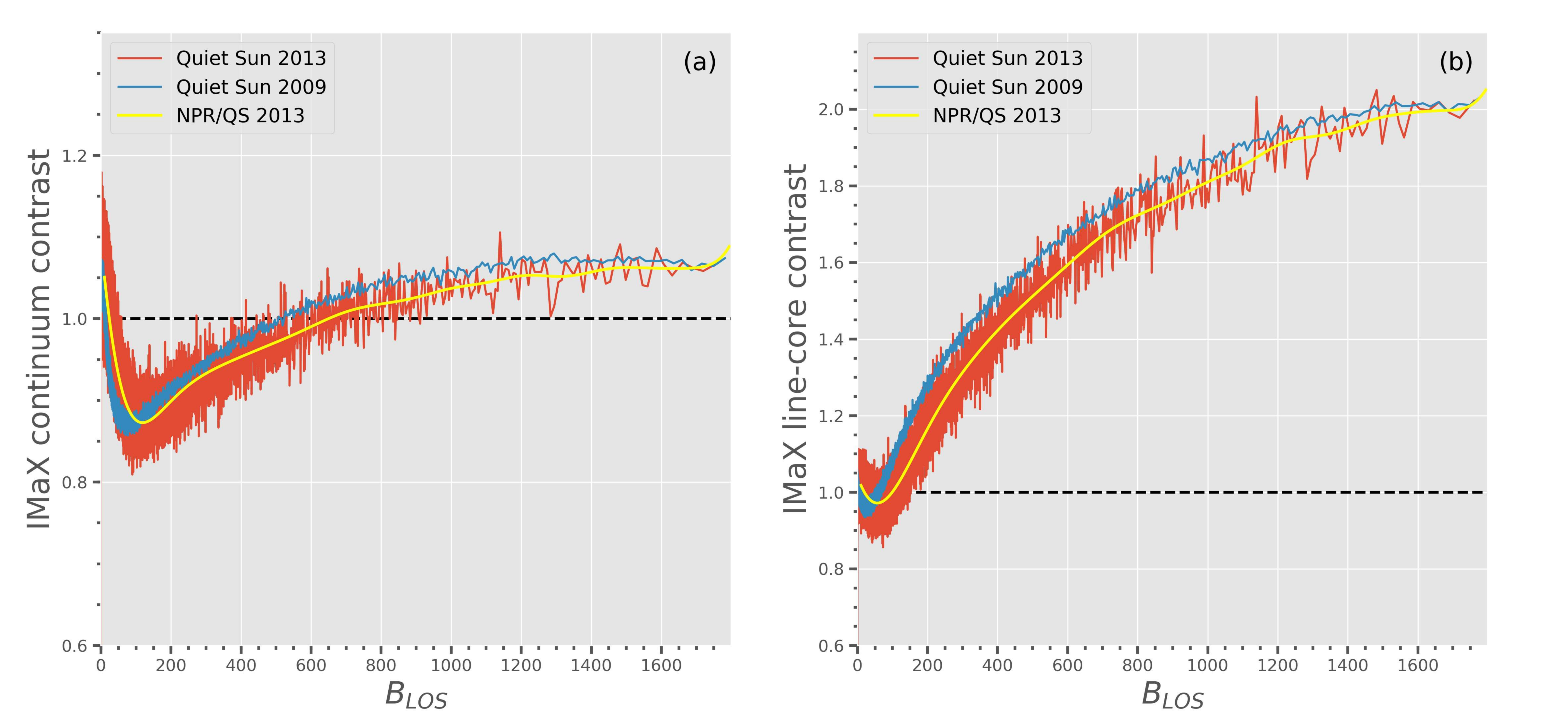}
\caption{IMaX continuum (panel a) and line-core (panel b) contrasts vs. $B_{\rm LOS}$. The red curves are the binned contrasts of quiet-Sun areas extracted from the 2013 ARs scans, and the blue curves are taken from Paper~I for quiet-Sun observations in 2009. The yellow curves are the result of non-parametric smoothing applied to the scattered data points of quiet-Sun pixels of 2013. The black dashed line is the quiet-Sun level where contrast is equal to unity.}
\label{qs_2009_2013}
\end{figure*}

In Figure \ref{qs_2009_2013} we show the scatterplots of the binned IMaX continuum (panel (a)) and line-core (panel (b)) contrasts, in blue for QS-2009 and in red for QS-2013, vs. $B_{\rm LOS}$. The black dashed horizontal line is the mean QS intensity level. Due to the smaller number of data points in QS-2013, the corresponding curves of averaged contrasts look noisier. Hence, we use non-parametric smoothing on the scattered data points of QS-2013 to make the trend easy for the eye to follow. The smoothed curves are overplotted in yellow in Figure~\ref{qs_2009_2013}.

According to Fig.~\ref{qs_2009_2013}, contrast-$B_{\rm LOS}$ relationships in both, IMaX continuum and line core agree qualitatively in the quiet-Sun regions from both flights.
In Paper~I we came to the conclusion that magnetic fields in the quiet-Sun network at disk center are resolved by IMaX since the contrast in the visible continuum saturated at higher $B_{\rm LOS}$. The fact that the quiet-Sun features embedded in AR scans exhibit the same behaviour implies that such fields are also resolved in $\sunrise$ II observations of active regions.
Not surprisingly, the behaviour of the averaged contrast variation with $B_{\rm LOS}$ of magnetic elements in the quiet Sun is independent of solar magnetic activity level \citep{ortiz_intensity_2006}, so that any difference between the two flights would reflect instrumental differences.

Quantitatively, in panel (a) of Figure~\ref{qs_2009_2013}, the minumum contrast of the fishhook in QS-2013 occurs at around 150\,G, a value higer than the 80\,G obtained for QS-2009. In addition, the features in QS-2013 start to become brighter than the mean QS at 650\,G, i.e. at fields higher than the 500\,G for QS-2009. \\ 
The averaged contrast of magnetic features in the QS-2009 data is higher than in the QS-2013 in both wavelength bands. Considering the basic picture of the `hot wall' and that observations in 2013 were done further from disk center ($\mu=0.93$) than in 2009 ($\mu = 0.98$), we expect that magnetic elements are brighter (in the visible continuum) with increasing heliocentric angle, but the opposite behaviour is seen in Figure~\ref{qs_2009_2013}a. However, the difference between the two data sets is small. The line core of IMaX (Figure~\ref{qs_2009_2013}b) displays the same behavior, which in this case is consistent with the findings of \cite{yeo_intensity_2013}.
The differences in the contrast could be interpeted as the effect of the higher average magnetic flux density in AR scans of 2013 data (50\,G) compared to the quiet Sun observed by $\sunrise$ in 2009 (30\,G). The convection is more hampered during the period of higher activity (2013), which leads to less effective heating inside the magnetic elements from their surroundings, and therefore less radiative energy emitted by these elements. 
\section{Discussion and conclusions}
\label{conclusion}
We have analyzed the properties of solar AR plage near disk center, as observed by $\sunrise$ during its second flight on June 2013. In particular, we have qualitatively and quantitatively described how the brightness in the visible and UV of the different components of magnetic features depends on the longitudinal field strength, $B_{\rm LOS}$. The latter is computed among other physical parameters from SPINOR inversions of IMaX Stokes profiles corrected for wavefront aberrations and straylight.

At the spatial resolution of 0.15$^{\prime\prime}$ achieved by IMaX, the scatterplot of the continuum contrast at 525 nm vs. $B_{\rm LOS}$ (Fig.~\ref{imax_cont_sp}) shows that the contrast peaks at around 850\,G beyond which it decreases with increasing $B_{\rm LOS}$, a behaviour that agrees qualitatively with the findings of \cite{berger_contrast_2007}, \cite{narayan_small-scale_2010} and \cite{kobel_continuum_2011}. The findings of the latter citation were interpreted to be an effect of the finite spatial resolution of the data \citep{danilovic_relation_2013}.

After locating the pixels in the $B_{\rm LOS}$ range of 600\,G-2000\,G, we found that these pixels belong to the same features, which are well resolved by $\sunrise$ (see Fig.~\ref{contour}). The $B_{\rm LOS}$ is always high in their centers and decreases towards their edges, while the contrast displays a more complex behaviour that can be different from one feature to another, depending on its size. In most of the features, however, the contrast is >1 in the limb direction, <1 towards disk center, and neutral near their centers (Fig.~\ref{profiles}d), whereas for larger size features, the centers are dark and brighten towards their edges (Figs.~\ref{profiles}a, b). These complex profiles are attributed to the fact that plage features are large and they are located slightly off-disk-center in the analyzed data \citep{steiner_recent_2005}.

Consequently, one and the same plage magnetic element provides pixels in different parts of the contrast vs. $B_{\rm LOS}$ plot, including bright/dark intermediate field pixels, and dark strong-field pixels. As a result, the turnover of brightness with $B_{\rm LOS}$ is not an effect of image smearing induced by the limited spatial resolution but rather of the internal morphology of individual plage magnetic features. This conclusion is strengthened by the saturation of the continuun contrast at higher $B_{\rm LOS}$ in quiet-Sun magnetic features (Paper~I and Sect.~\ref{QS_2009_2013}), which are smaller than the features in the plage, so that their contrasts are more likely to suffer from insufficient spatial resolution. The network features also show much less internal structure.

We think that the simulations used so far to interpret the downturn of the contrast in observations as an artifact of the poor spatial resolution \citep{rohrbein_is_2011, danilovic_relation_2013}, lack the larger magnetic features found in strong plage that dominate their continuum contrasts. By introducing higher average magnetic flux ($\approx$ 400\,G) and by extending the vertical depth, we expect the boxes to be deep enough to allow convection to produce larger features. In addition, the current simulations for plage regions are analyzed for lines-of-sight parallel to the solar normal (i.e. $\mu=1$). Therefore, we propose that for such simulations to reproduce our data, bigger and deeper simulation boxes are needed, in addition to performing spectral line synthesis with such simulations for lines-of-sight that are inclined to the local normal ($\mu=0.93$).

In the UV at 397\,nm (core of the chromospheric Ca\,{\sc ii} H line) and in the core of the IMaX line, the averaged contrasts of plage features increase with increasing $B_{\rm LOS}$. Given the larger formation heights of these wavelengths compared to the IMaX continuum, this implies that the brightness at larger atmospheric heights is independent of the feature size, while lower in the photosphere, the size of the magnetic features plays a decisive role in determining their contrast \citep{solanki_small-scale_2001}.  

Comparison of the contrast-$B_{\rm LOS}$ relationship in plage with quiet-Sun observations presented in Paper~I confirms the findings of earlier studies: at the photospheric level and for larger field strengths, the contrast in the quiet Sun network is higher than active region plages \citep{title_differences_1992, morinaga_suppression_2008, kobel_continuum_2011} due to the less efficient convective energy transport in the latter \citep{morinaga_suppression_2008, kobel_continuum_2012,criscuoli_comparison_2013}.

In the cores of spectral lines, this difference is found to be smaller. At the atmospheric layers sampled by these wavelengths (upper photosphere and lower chromosphere), radiative heating from convection becomes less efficient in determining the contrast of these features, and other processes dominate in transporting and dissipating energy (e.g. oscillations and waves). 

\begin{acknowledgements}

The German contribution to \sunrise{} and its reflight was funded by the
Max Planck Foundation, the Strategic Innovations Fund of the President of the
Max Planck Society (MPG), DLR, and private donations by supporting members of
the Max Planck Society, which is gratefully acknowledged. The Spanish
contribution was funded by the Ministerio de Econom\'{\i}a y Competitividad under
Projects ESP2013-47349-C6 and ESP2014-56169-C6, partially using European FEDER
funds. The HAO contribution was partly funded through NASA grant number
NNX13AE95G. This work was partly supported by the BK21 plus program through
the National Research Foundation (NRF) funded by the Ministry of Education of
Korea.
\end{acknowledgements}

\bibliographystyle{aa}
\bibliography{kahil_ms_2} 

\begin{appendix}

\section{The quiet Sun's mean intensity in SuFI data}
\label{appendix1}

We use the dark corrected (level~$0.1$) data taken over nearly 4 hours (from June 12, 22:15:37 UT to June 13, 01:58:00 UT) to visualize the day-to-night cycles of the mean quiet-Sun intensity at 300\,nm and 397\,nm. Quiet Sun images are acquired for more than an hour at the beginning of the series, just prior to the 1\,h spent observing the active region, and in three intervals after the end of the AR observations. 

\begin{figure*}[ht!]
\begin{center}
\includegraphics[scale=0.56]{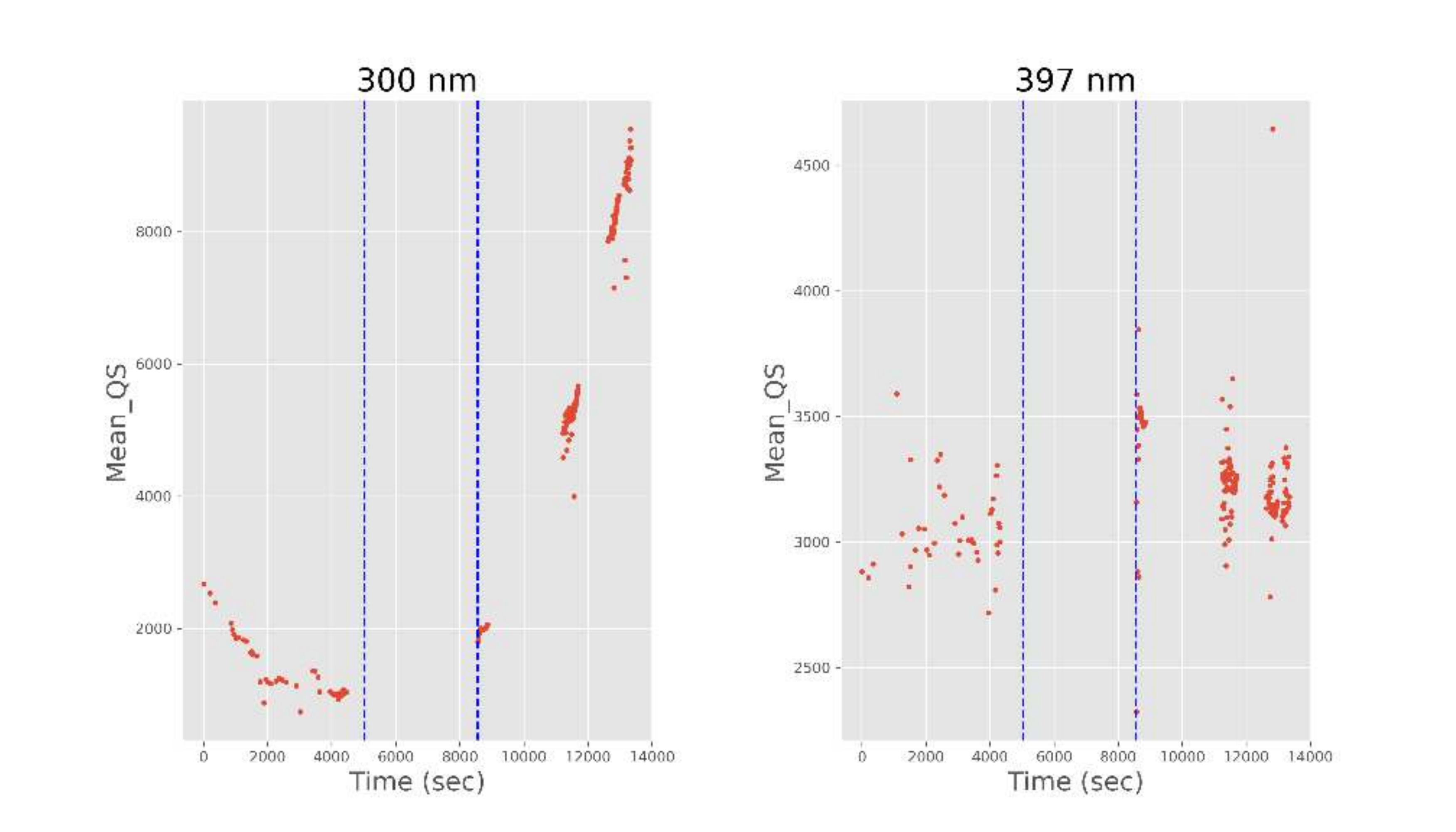}
\caption{Part of the day-to-night cycle of the photon flux at 300\,nm and 397\,nm for quiet-Sun images taken at disk center. The dashed blue lines delimit the 1 hour observation of the active region, which was partly analysed in this work.
}
\end{center}
\label{cycle}
\end{figure*}

We plot for both wavelengths the averaged level~$0.1$ QS intensity ($I$) versus time ($t$), with the time at which the first image of the time series was recorded being referred to as $t=0$. The day-night cycle is clearly visible in the left panel of Figure~\ref{cycle}. 

For fitting the day-to-night cycles we use the Beer-Lambert law, which gives the amount of flux (of original value $I_0$) absorbed by a medium with a Rayleigh airmass of $m$: 

\begin{equation}
I = I_0  \times e^{- m\tau_{\lambda}}.
\label{beer}
\end{equation}

$\tau_{\lambda}$ is the optical depth of the terrestrial atmosphere at a given wavelength $\lambda$. The Rayleigh airmass depends on the Sun's elevation angle, $\phi$ \citep{pickering_southern_2002}, and for a spherically-symmetric atmosphere it is given by:

\begin{equation}
m = \frac{1}{sin(\phi +\frac{244}{165 + 47\phi^{1.1}})}.
\end{equation} 

\begin{table}[ht!]
\caption{Best-fit parameters at each wavelength obtained by using Eq.~\ref{beer} and applied to $I$ vs.~$m$ scatterplots. }
\label{fit_params}
\centering
\begin{tabular}{c c c}
\hline \hline

Wavelength (nm) &$I_0$ & $\tau_{\lambda}$  \\
\hline
300 &  188579.7 & 0.3 \\
\hline
397 & 3198.101 &  -0.0003 \\
\hline
\hline
\end{tabular}

\end{table}

\begin{figure*}[ht!]
\centering
\includegraphics[scale=0.1]{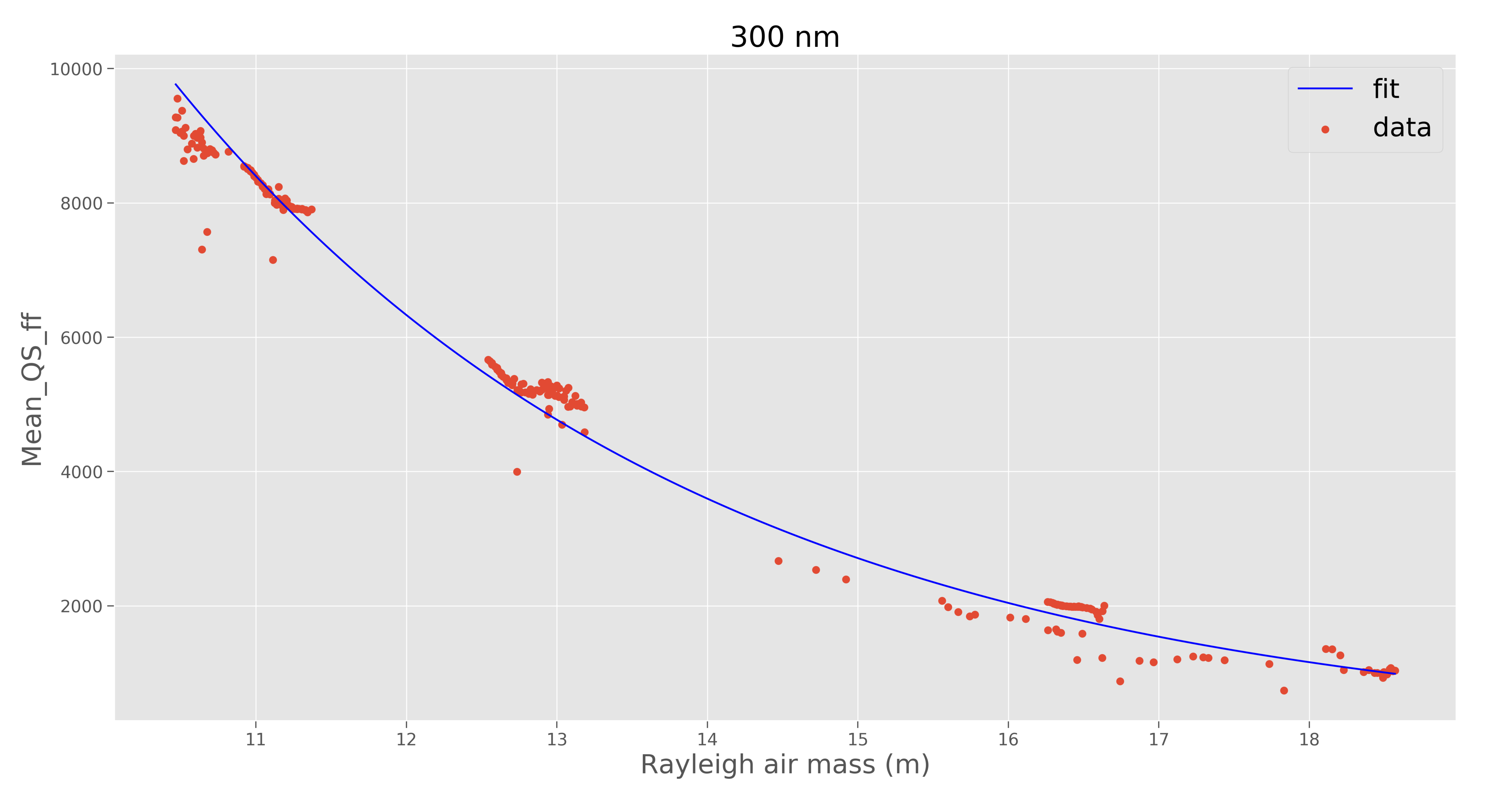}
\caption{Red data points: the 300\,nm quiet-Sun photon flux vs. $m$, the air mass factor during the 4 hours observing period. The best-fit blue curve was calculated according to Equation~\ref{beer} with the parameters given in Table~A1.}
\label{I_m_300}
\end{figure*}

\begin{figure*}[ht!]
\centering
\includegraphics[scale=0.1]{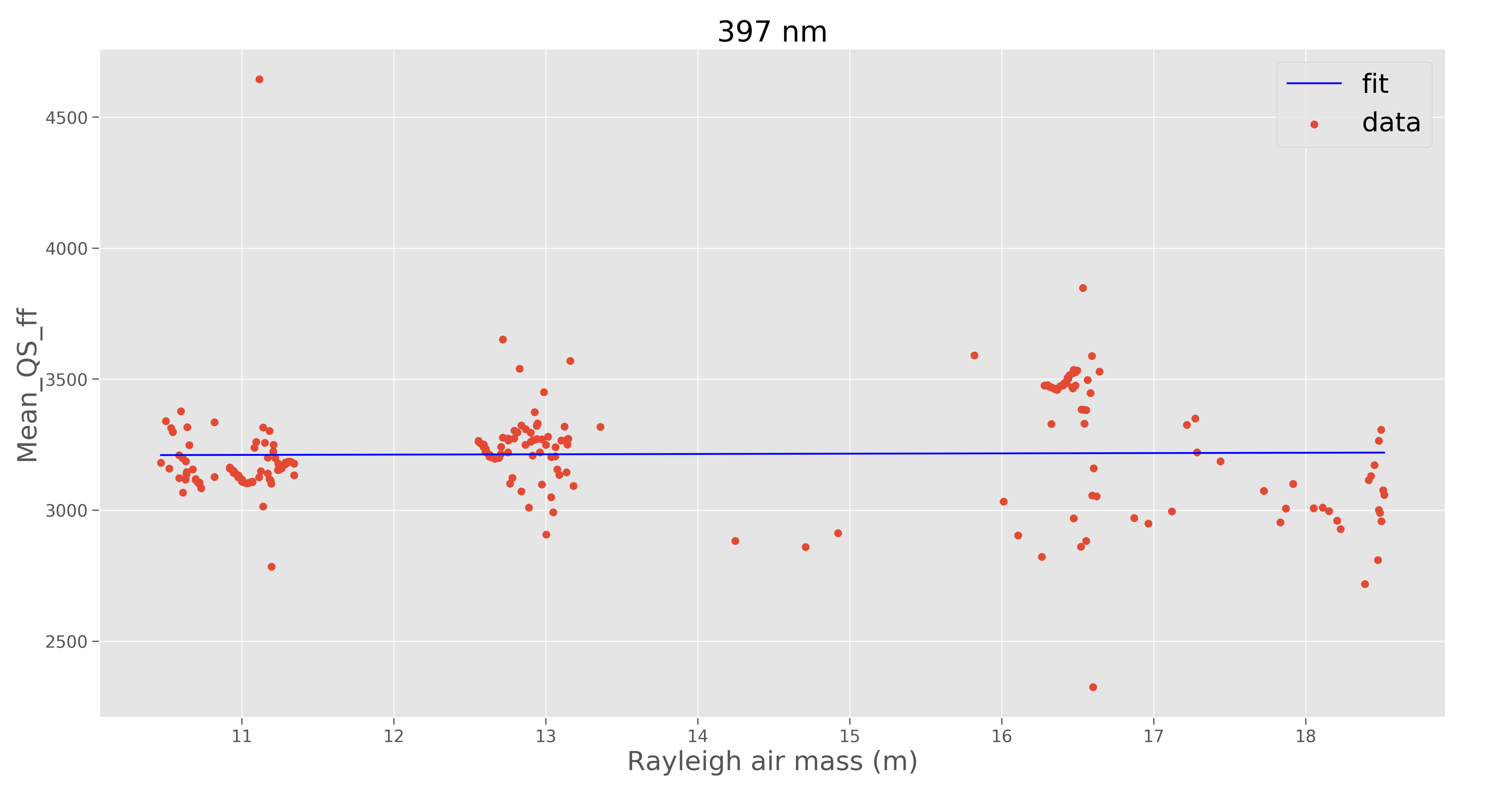}
\caption{Same as Figure~\ref{I_m_300}, but for the 397\,nm data.}
\label{I_m_397}
\end{figure*}

For each wavelength, $I_0$ and $\tau_{\lambda}$ are determined from a fit to the $I$ vs.~$m$ data based on Eq.~\ref{beer}. They are shown in Table~A1. Plots of $I$ versus $m$ and the corresponding fits are shown in Figures~\ref{I_m_300} and \ref{I_m_397} for 300\,nm and\,397 nm, respectively.
As expected, the day-to-night cycle variation is negligible at 397 nm since the average quiet-Sun intensity does not vary with time (or elevation angle). Therefore, the average of the flatfields is the average QS intensity. 

At 300\,nm, we use the best-fit parameters to evaluate $I$ at the airmass values of our AR data. The $I$ values determined in this way are the true mean quiet-Sun intensities. We normalize each image in our level 3.1 data with the corresponding evaluated mean after restoring the original flux. The latter is simply the product of every level~$3.1$ pixel value with the averaged flatfield to which our data were normalized (assuming that the straylight-correction and phase-diversity reconstruction do not affect the mean value of the image).\\

\end{appendix}

\end{document}